\documentclass[final,onecolumn,12pt]{article}
\usepackage{amssymb}

\usepackage{graphicx}
\usepackage{amsmath}
\usepackage{citesort}

\input{tcilatex}

\begin{document}

\title{Thermodynamics and Structure of Self-assembled Networks}
\author{A.G. Zilman\thanks{%
e-mail:anton.zilman@weizmann.ac.il} \ and S.A. Safran \and \textit{Dept. of
Materials and Interfaces} \and \textit{Weizmann Institute of Science} \and 
\textit{76100 Rehovot, Israel}}
\maketitle

\begin{abstract}
We study a generic model of self-assembling chains which can branch and form
networks with branching points (junctions) of arbitrary functionality. The
physical realizations include physical gels, wormlike micells, dipolar
fluids and microemulsions. The model maps the partition function of a
solution of branched, self-assembling, mutually avoiding clusters onto that
of a Heisenberg magnet in the mathematical limit of zero spin components.
Regarding thermodynamics properties as well as scattering strucuture factor,
the mapping accounts rigorously for all possible cluster configuration,
except for closed rings. The model is solved in the mean field
approximation. It is found that despite the absence of any \textit{specific
interaction} between the chains, the presence of the junctions induces an 
\textit{effective attraction} between the monomers, which in the case of
three-fold junctions leads to a first order reentrant phase separation
between a dilute phase consisting mainly of single chains, and a dense
network, or two network phases. The model is then modified in order to
predict the structural properties on the mean field level. Independent of
the phase separation, we predict the percolation (connectivity) transition
at which an infinite network is formed. The percolation transition partially
overlaps with the first-order transition. The percolation transition is a
continuous, \textit{non thermodynamic }transition that describes a change in
the topology of the system but not a thermodynamic phase transition. Our
treatment which predicts both the thermodynamic phase equilibria as well as
the spatial correlations in the system allows us to treat both the phase
separation and the percolation threshold within the same framework. The
density-density correlation correlation has a usual Ornstein-Zernicke form
at low monomer densities. At higher densities, a peak emerges in the
structure factor, signifying an onset of medium-range order in the system.
Implications of the results for different physical systems are discussed.
\end{abstract}

\section{Introduction\label{intro}}

Networks and branched structures are ubiquitous in both natural and
synthetic materials and form under a variety of equilibrium and
non-equilibrium conditions. In this paper we present a theory that describes
in a unified way the structure and thermodynamic properties of equilibrium
networks and their relation to several soft condensed matter systems, such
as gels, worm-like micelles, microemulsions, and dipolar liquids and
colloids. In all these systems the networks consist of cross-linked
elongated objects (e.g., polymer chains in the case of a gel). The study of
network phases is of both theoretical and practical interest. From the
practical point of view, gels and sol-gel systems are at the core of many
industrial, biological and bio-medical applications. Examples range from
novel plastics and food processing to gel chromatography and tissue implants 
\cite{mrs,kumar}. Despite intensive theoretical and experimental
investigations carried out over the past three decades, many features of
network-forming systems still remain unclear. For example, whether the
gelation transition in the physical gels is a first-order or a continuous
transition, is currently under debate \cite
{tanakatheor,rubinstein,physgel1,kumar}. Experiments have not yet provided a
clear-cut answer, because the gelation transition can be obscured by the van
der Waals interaction \cite{tanakaexp}. Another example is the transition of
a solution of worm-like micelles to a self-assembled network that has been
suggested\ to occur on the basis of rheological measurements; this has also
been discussed theoretically \cite{khatory,lekek}. Network formation may
also be responsible for the `closed loop' phase diagrams of dipolar and
magnetic liquids and colloids and of microemulsions \cite{tsvidip}.
Cryo-electron microscopy shows clear evidence of coexisting network phases
in dilute microemulsions\cite{talmon}. In many of the systems mentioned
above, the chains themselves are self-assembled in an equilibrium manner
from a large number of monomers.

We focus here on the thermodynamic behavior and structure of \ systems with
thermoreversible cross-linking; this means that the cross-links can break
and reform under the influence of thermal fluctuations. It turns out
however, that many of the large scale structural properties of
network-forming phases are independent of the precise nature of the
cross-links. In many of the complex systems mentioned above, an
understanding of the local and large scale correlations are also of
importance. Certain structural transitions cannot be detected by examining
the thermodynamic properties alone, but are expressed by the nature of the
correlations between different components of the system. One example is the
percolation transition, at which a network spanning the whole system is
formed, as schematically shown in Fig. \ref{percfig}. It is a continuous
transition, unrelated to the thermodynamic properties of the system.
Generally speaking, one would like to know both the thermodynamic behavior
of these systems as well as such experimentally measurable quantities as
density-density correlation functions and response functions to applied
fields such as an electromagnetic field or hydrodynamic flow.

To answer these questions, at least within a mean-field approach,\emph{\ }we
study a generic system consisting of self-assembled chains which can branch
and form networks. Each chain consists of a large number of \ `monomers'.
The chains are self- and mutually- avoiding but are permitted to branch
(cross-link). We shall call both the branching points and the cross-links
`junctions'. The exact physical interpretation of the `monomers' and
`junctions' will differ form system to system. In each particular case, the
energy and other properties of the junctions and the ends can be calculated
from microscopic considerations, e.g., molecular packing for surfactant
systems, or dipolar interaction energies applicable to ferrofluids.

We present here a lattice model which establishes the equivalence between a
solution of \ branched self-avoiding self-assembling chains and a Heisenberg
magnet in the limit of `zero' spin components; we show that their partition
functions can be mapped onto another. This model, known as `$n=0$' model,
was proposed by De Gennes to study polymer solutions and we have modified it
in order to include the possibility of branching. Regarding the
thermodynamic properties, including the density fluctuations, the model is
exact and accounts for all possible configurations of equilibrium branched
clusters (excluding separate closed rings, whose influence is or importance
only at very low densities). This formulation of the problem enables us to
explore both thermodynamic and structural properties of self-assembling
branching chains in a unified manner. The model has been treated in the mean
field approximation and our major conclusions are summarized in Fig.\emph{\ }%
\ref{fig1}, which shows the phase diagram of the system as a function of the
density and temperature. We find that despite the \textit{absence of any
specific interactions} between the monomers, the presence of the junctions
induces an \textit{effective attraction} between the monomers. The model can
be amended in order to study the topological properties of the system and
correlate them with the thermodynamic behavior. As the monomer density is
increased or the temperature decreased, the system passes through a \textit{%
percolation (connectivity)} threshold, where\ a network spanning the entire
volume is formed. This transition is purely \textit{topological,} and has no
thermodynamic signature. However, the junction-induced attraction does
modify the concentration of monomers at which the percolation threshold
occurs; as the temperature is decreased, the threshold is decreased compared
with its value in the limit of infinite temperature, where the interactions
are irrelevant.

For three-fold junctions, the junction-induced attraction is strong enough
to drive a $\ $\textit{first-order} \textit{phase separation}, where the
system separates into a low-density and a high density phases. This
transition is of purely entropic origin, because there are no any specific
interactions between the monomers. The physical reason for the transition is
the higher entropy of the junctions-dominated high density phase: although
the translational entropy of the chains is lower in the high density phase,
it is overcompensated by the increase of the entropy of the self-assembled
junctions, abundant in the high-density phase. In this respect, this
entropy-induced transition is similar to the crystallization of \ rigid
spheres and the isotropic-nematic transition. The transition line terminates
at a critical point. For four-fold and higher functionality junctions, the
junction induced transition is too weak to drive \ a phase separation. In
this case, the junction-induced interaction merely\ \ renormalizes the
excluded volume interaction between the chains, and drives the system closer
to the $\Theta $-point. The structural percolation transition, of course, is
also present above the critical temperature of the first-order phase
separation. At very low temperatures and monomer densities, the chains are
depolymerized and the system consists mainly of separate monomers.

The nature of the phases in equilibrium depends on the temperature, but also
on the rigidity of the chains and details of the junction configurations. In
general, there are three possibilities:\textit{\ (i)} a phase of dilute
chains that coexists with a \textit{connected network}, \textit{(ii)} two
coexisting networks, or \textit{(iii)} the coexistence of dilute and dense
phases of\ disjointed, branched aggregates, although not predicted by our
model, is also possible.

Treatment of the spatial density variations within the mean field theory
shows that the density-density correlation function, relevant for scattering
experiments, has a simple Ornstein-Zernicke form for relatively low
densities of the self-assembling monomers. For higher densities, a \textit{%
peak} emerges in the structure factor as a function of the scattering
wavevector; this indicates the emergence of medium range (longer range than
the lattice size but not long range order)\ correlations in the system. In
the intermediate regime the structure factor is a monotonically increasing
function of the wavevector. This is expected in any dense system: at high
densities, the system has a low compressibility at long wavelengths and this
suppresses the small wavevector scattering. The \textit{absence of a peak} $%
\ $in\ the structure factor does not imply the absence of the\textit{\
network}: there is a region in the phase diagram beyond the percolation line
( where an infinite network exists) where there is indeed no peak in the
density-density correlation function. The precise location and strength of
the peak depends on the monomer density and the number of junctions present.
As discussed in Section. \ref{discussion}, these theoretical predictions may
be related to the scattering peak observed experimentally in bicontinuous
microemulsions.

The predictions of the thermodynamic and structural properties can be put in
the context of previous work on the physics of `living' (self-assembling)
polymers and networks; this area has been the focus of extensive
experimental and theoretical attention over the past two decades. Most
notably, phenomenological Flory type (chains with linkers) theories have
been employed to model physical gels \cite{rubinstein,tanakatheor}. Another
important contribution is the work of Drye and Cates that was motivated by
studies of worm like micelles \cite{drye}. However, the predictions of these
various studies regarding the properties of the gelation transition vary
depending on the details of the models that are studied, the assumptions
regarding the solvent quality and other parameters. This is partially due to
the fact that these studies \cite{rubinstein,tanakatheor} did not emphasize
the extreme sensitivity\ of the gelation transition to the functionality of
the junctions and the importance of \emph{the competition of the branching
points and} \textit{the free ends}. These previous, more heuristic,
mean-field theories could not be extended in a simple manner to predict the
correlations.

We have therefore focused on the more rigorous `$n=0$' model, which allows
one to evaluate \textit{exactly} the number of all possible configurations
of branched self- and mutually avoiding chains on a lattice. Lattice
approximation models the configurations of the chains only approximately at
short lenghtscales. However, as long as one is interested the behavior of
the system on lengthscales much larger than the lattice constant, (i.e.,
thermodynamics) the choice of the lattice does not influence any of the
qualitative predictions of the model. The '$n=0$' model has been
successfully used to explain the properties of associating monomers and has
been applied to the polymerization of sulfur \cite{wheelersulf} and worm
like micellar solutions \cite{Gelbartn=0}. The extension of the model to $%
n=1 $ was used to model systems consisting of chains and rings in
equilibrium \cite{petschek}. Our model is similar to the one employed by
Isaacson an Lubensky \cite{isaacson} to which it can be related (cf.
Appendix D). These authors studied the continuum version of the `$n=0$' \
model with a cubic term in the Hamiltonian in the context of percolation
focusing on the scaling behavior of the model to study the properties of the
gelation-percolation transition but did not calculate the thermodynamic
transitions (such as the phase separation predicted here) or \ study the
monomer density fluctuations\ and scattering structure factor. In our model,
however, the junctions are not modelled as point-like objects, and the
treatment of topological structure is simpler than of Ref. \cite{isaacson},
but is applicable only in the mean field approximation. The $n=1$ model,
which accounts also for closed rings, but does not properly take into
account the self-avoidance, was used by Lequeux, Elleuch and Pfeuty \cite
{lekek} to study general properties of micellar solutions without
considering the topology of the network. Similarly, in the two-field model
of Ref.\cite{panizza}, correlations and structure were not studied and no
analytic expression for the free energy was obtained.

This paper is organized as follows. In Sec. \ref{chainsn0} we review the `$%
n=0$' model without junctions. In Sec. \ref{juncn0} we extend the model to
include the possibility of junctions and establish the correspondence
between this model and the physics of the problem of self-assembled
networks. In Sec. \ref{mf} we calculate the free energy and phase diagram in
the mean-field approximation. We next extend the mean field theory to
include spatial variations of the monomer density and in Sec.\ref{corr}%
,predict the spatial variations and correlations. In the Sec. \ref{spinspin}
we show how the model can be modified on the mean-field level to enable us
to predict the topological properties and correlate them with the
thermodynamic behavior. The approximation employed, is of the same level as
the classical Flory-Stockmayer theories of gelation. In particular, in Sec. 
\ref{transversal} we discuss the structure and evolution of a single
branched cluster as a function of monomer density and the temperature, and
the emergence of a connected network. Sec. \ref{rigid} describes the
extension of our model and its predictions to the case of rigid and
semiflexible chains. In Sec. \ref{discussion} we discuss the results and
their application to particular physical systems\ such as physical gels,
microemulsions, dipolar fluids and wormlike micelles.

\FRAME{ftbpFU}{3.4108in}{2.5797in}{0pt}{\Qcb{Phase diagram of solution of
self-assembling branched chains as a function of temperature $T$ in units of
the end energy $\protect\epsilon _{e}$, and monomer volume frcation $\protect%
\phi .$ \ The junction energy has been chosen to be $\protect\epsilon _{j}=%
\protect\epsilon _{e}/4$. The thick line shows the phase separation region.
The dashed line is the percolation line. To the left of this line the system
consisits a solution of disconnected (but possibly entangled) branched
chains; to the right there is an infinite connected network. Moving along
the line $b$, the actual gelation transition will be seen as a continuous
non-thermodynamic transition, while moving along $a$ will result in the
first-order thermodynamic transition. Thus, coexistence of two connected
networks is possible at high enough temperatures.}}{\Qlb{fig1}}{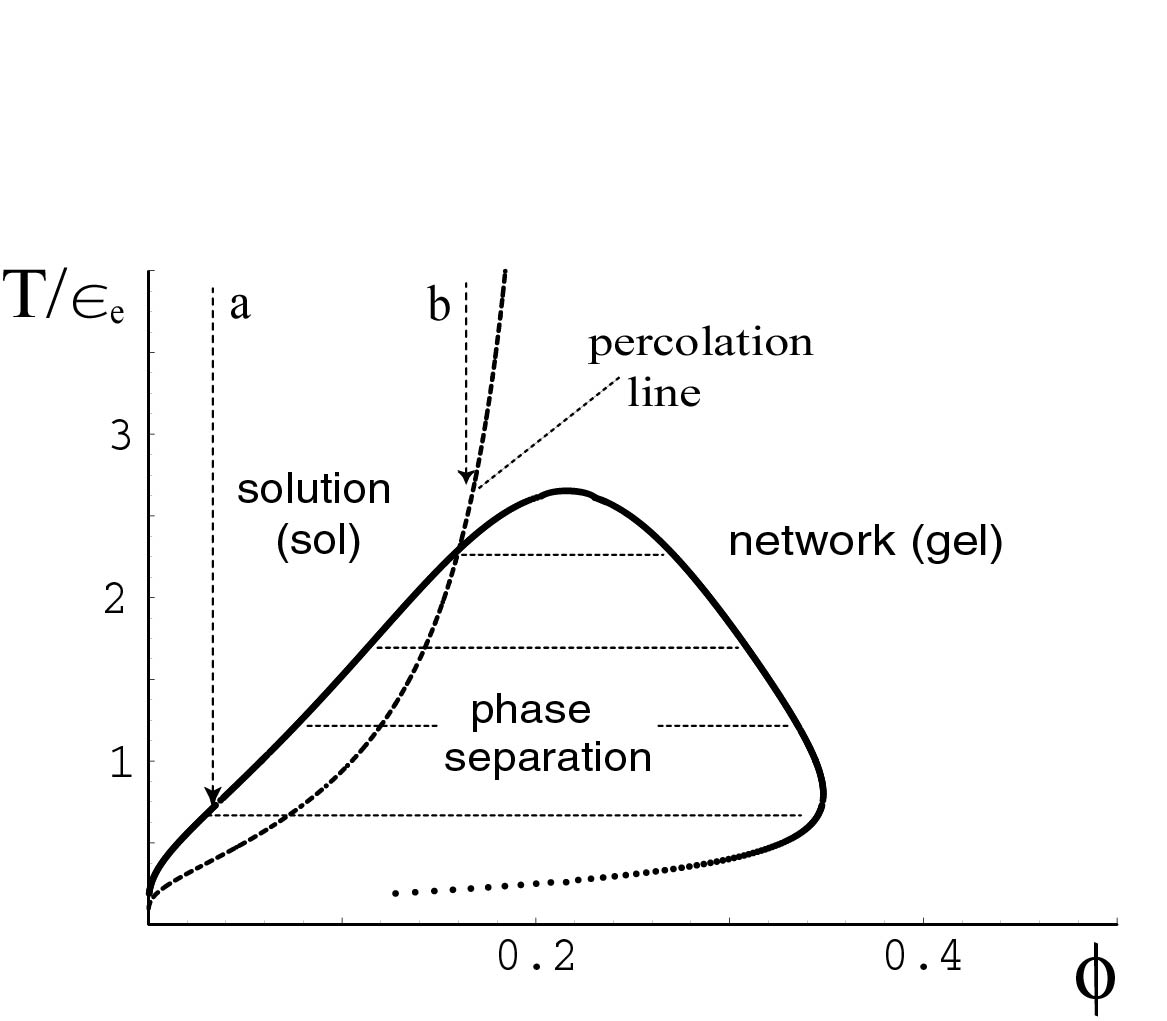%
}{\special{language "Scientific Word";type "GRAPHIC";display
"USEDEF";valid_file "F";width 3.4108in;height 2.5797in;depth
0pt;original-width 24.3125in;original-height 21.2709in;cropleft "0";croptop
"1";cropright "1";cropbottom "0";filename
'L:/networks/phd1flex.jpg';file-properties "XNPEU";}}

\section{Zero-component' Heisenberg model\label{n=0}}

\subsection{Self-avoiding chains.$\label{chainsn0}$\qquad}

For systems with no junctions, the `$n=0$' model has been extensively
studied in the context of polymer solutions and micellar systems \cite
{doibook,sarma,wheeler,Gelbartn=0}. For the sake of completeness we review
here the derivation of the model; in Sec. \ref{juncn0}\ we\emph{\ }extend
the model to include junctions. Readers familiar with `$n=0$' model can
proceed directly to Sec. \ref{juncn0}. We consider a system that comprises a
solvent and a collection of self-assembling chains, each consisting of a
large number of identical monomer units. The physical nature of the
`monomers' can be different. For example, in physical gels these would be
the individual molecular units that comprise the polymers, in microemulsions
these would be surfactant covered oil (or water) domains and in dipolar
colloids, the colloidal particles (cf. the discussion in Sec. \ref
{discussion}). The chains are self- and mutually avoiding, that is, they
cannot intersect themselves and each other. We focus on self-assembling
systems where the monomers are relatively weakly associated within a chain;
thus, individual monomers can be freely exchanged between the chains. In
this case, the equilibrium distribution of chain lengths is not fixed, but
is determined by the values of relevant physical parameters, as we discuss
later. We will consider the system in the grand-canonical ensemble, in
equilibrium with the bath of monomers of the chemical potential $\mu $. We
assume that the energy of a monomer at the end of the chain, $\epsilon _{e},$
relative to that of a monomer in the middle of the chain is positive, $%
\epsilon _{e}>0$. This is indeed the case for micellar systems and
microemulsions, because (in the regime where long chains are formed) the
bending energy of the surfactant layer is lower in the cylindrical part than
in the semi-spherical end-caps \cite{sambook}. The same is true for dipolar
particles, because the electrostatic energy of a dipole in the middle of the
chain with two neighbors, is lower than that of a dipole at the chain end%
\emph{, }where it has only one neighbor dipole. If $\epsilon _{e}<0$, \emph{%
\ }the self-assembling chains will be very short and the solution will
mostly consist of individual monomers; this is not the limit that we are
interested in. \ The end energy, taken with the opposite sign, \emph{-}$%
\epsilon _{e},$ can also be interpreted as $\mu _{p}$, the chemical
potential conjugate to the number of chains in the solution, $N_{p}$. This
interpretation is usually adopted in discussion of regular polymer solutions 
\cite{doibook,sarma}. For the systems of interest, the energy $\epsilon _{e}$
is determined from microscopic considerations, be it molecular packing or
electrostatics. The number of chains and average molecular weight are not
fixed but determined by physically controllable parameters, namely, the
temperature, the total \textit{monomer} density $\phi $ and $\epsilon _{e}$.
For simplicity we imagine that the monomers and the solvent molecules occupy
the sites of a $d$-dimensional lattice. With the above notation, the grand
canonical partition function of a solution of equilibrium chains is given by 
\begin{equation}
Z=\sum_{\{N,N_{e}\}}\exp [\mu N/T]\exp \left[ -\epsilon _{e}N_{e}/T\right] 
\mathcal{N}(N,N_{e})  \label{Z}
\end{equation}
where $N$ is the number of monomers in a given realization, $N_{e}$ is the
total number of free ends, and $\mathcal{N}(N,N_{e})$ is the number of ways
to arrange $N_{e}/2$ chains of total length $N$ on a lattice. Thermodynamic
quantities can be obtained from Eq.$\left( \ref{Z}\right) $ by
differentiation with respect to the relevant parameter. For example, the
mean number of monomers and chains, $\bar{N}$ and $\bar{N}_{p}$
respectively, are:

\begin{eqnarray*}
\bar{N} &=&\frac{\partial \ln Z}{\partial \mu } \\
\bar{N}_{p} &=&\frac{\partial \ln Z}{\partial \epsilon _{e}}
\end{eqnarray*}

The direct combinatorial calculation of $\mathcal{N}(N,N_{e})$ is a
formidable task. A way of circumventing the difficulties involved in its
calculation, was proposed by De Gennes \cite{degen1972}. It was later
refined by equivalent methods by different authors \cite{descloiseaux,witten}%
. The calculation of $\mathcal{N}(N,N_{e})$ is done by relating the
partition function of the Eq.(\ref{Z}) to that of a certain spin model on a
lattice. Consider the\emph{\ }$n$-component Heisenberg model on a $d$%
-dimensional lattice whose Hamiltonian is given by 
\begin{equation}
H=-\sum_{\{\vec{S}_{i}\}}J\vec{S}_{i}\cdot \vec{S}_{j}-\sum_{\{\vec{S}_{j}\}}%
\vec{h}_{i}\cdot \vec{S}_{i}  \label{H}
\end{equation}
where the summation is over all $distinct$ pairs\ of nearest neighbors $i,j$%
. Spin $\vec{S}_{i}$ is an $n$-component vector normalized at each site as 
\begin{equation*}
\left| \vec{S}_{i}\right| ^{2}=\sum_{\alpha =1}^{n}S_{i,\alpha }^{2}=n
\end{equation*}
where $\alpha $ indexes the components of $\vec{S}_{i}$. We arbitrarily
choose the field $\vec{h}$ to point in the direction $(1,0,...,0)$.

The partition function corresponding to this spin Hamiltonian is 
\begin{equation}
Z_{H}=\text{Tr }\exp [-H/T^{\star }]  \label{zh}
\end{equation}
\emph{\ }where the trace operator $Tr$ signifies an integration over all
possible directions of $\vec{S}_{i}$, divided by a normalization, so that
the trace $Tr\,A$ of any quantity\ $A$ is: 
\begin{equation*}
\text{Tr }A=\frac{\int \prod_{i}d\Omega _{i}^{(n)}A}{\int \prod_{i}d\Omega
_{i}^{\left( n\right) }}
\end{equation*}
where $\Omega ^{\left( n\right) }$ is the $n$-dimensional solid angle. The
`temperature' $T^{\star }$ has no physical meaning and is not related to the
real physical temperature of the system of in Eq.$\left( \ref{Z}\right) $;
we put $T^{\star }=1$ (another choice of $T^{\star }$ would amount to a
renormalization of the constants $J$ and $h)$. Expanding the exponential in $%
Z_{H}$ of Eq.$\left( \ref{zh}\right) $ in a power series, we obtain 
\begin{equation}
Z_{H}=\sum_{k}\frac{1}{k!}\text{Tr}\prod_{\langle i,j\rangle }(J\vec{S}_{i}%
\vec{S}_{j}+hS_{i,1})^{k}  \label{expansion}
\end{equation}
Each term in Eq.(\ref{expansion}) has the following form 
\begin{equation}
\text{Tr [}J^{m}\underset{m\text{ }times}{\underbrace{S_{i}S_{j}...S_{i^{%
\prime }}S_{j^{\prime }}}}h^{n}\underset{n\text{ }times}{\underbrace{%
S_{p,1}...S_{p^{\prime },1}}}\text{]}  \label{expansionterms}
\end{equation}
Note that in the tracing $operation$ in Eq.(\ref{expansionterms}), all sites
are decoupled.

In any realistic system in which the spin $\vec{S}$ with $\ n$ components is
to have a physical interpretation, $n\geq 1$ is required$;$ $n=1$
corresponds to the usual Ising model. However, quantities such as partition
function and the cumulant expansion remain mathematically meaningful even
when $n<1$. In particular, one can consider the mathematical limit $%
n\rightarrow 0$ \cite{degenbook,sarma}. It can be shown that when $%
n\rightarrow 0$ (cf. Appendix \ref{n=0app})$,$

\begin{equation}
\text{Tr }S_{i,\alpha }S_{j,\beta }=\delta _{\alpha \beta }\delta _{ij}
\label{n0}
\end{equation}
and \textit{all other cumulants} are zero. The spin $\vec{S}$ does not have
a physical interpretation in the $n\rightarrow 0$ limit and should be
regarded purely as a convenient mathematical device. Precise mathematical
meaning can be given to Eq.(\ref{n0}) (cf. Appendix \ref{n=0app}). In
particular, Eq.(\ref{n0}) means that in the expansion $\left( \ref
{expansionterms}\right) $ the only non-vanishing terms are those in which
every spin $S_{i}$ appears twice. As can be seen from Eq.$\left( \ref
{expansion}\right) ,$ this condition is satisfied by terms consisting of the
chains of neighboring bonds: $...JS_{i}S_{j}JS_{j}S_{j^{\prime }}...$. Since
every spin must appear twice for the term to be non-zero, the uncoupled $%
spins$ must be paired off with another spin at the same site either by \emph{%
(i)}\ closing a chain on itself or by \emph{(ii)}\ the appearance of the
single-spin, `field' term $h_{1}S_{1}$. However, in the limit of $%
n\rightarrow 0$, the terms given by\emph{\ (i)} vanish, due to summation
over all the components (via summation over the index $\alpha );$ this sum
is proportional to the number of components\emph{,} $n,$ which is equal to
zero. The terms that contain the field variables, $h_{1},$ do not vanish
because the $h$-terms single out $one$ component $\alpha ,$ parallel to the
field $h$. It is easy to see that such terms consist of products of the form 
\begin{equation}
\text{Tr }%
h_{i,1}S_{i,1}JS_{i,1}S_{j,1}JS_{j,1}S_{k,1}...JS_{q,1}S_{r,1}h_{j,1}S_{r,1}
\label{singlewalk}
\end{equation}
that repeat themselves with different $i$ and $r$. Any term of the form of
Eq.$\left( \ref{singlewalk}\right) $ corresponds to a self-avoiding random
walk on a lattice, starting at site $i$ and ending at site $r$. In other
words, a term containing the factor $J^{m}h^{2k}$ counts all the possible
configurations of $k$ random walks of total bond length $m$. Thus, the
partition function can be written: 
\begin{equation}
Z_{H}=\sum_{\{N_{b,}N_{e}\}}J^{N_{b}}h^{N_{e}}\mathcal{N}(N_{b},N_{e})
\label{zzh}
\end{equation}
where $\mathcal{N}(N,N_{e})$ is the number of ways to arrange on a lattice,
an ensemble of\ self-avoiding random walks with a total number of $bonds$ $%
N_{b}$ and a total number of ends $N_{e}$. Noting that the number of chains
is $N_{p}=N_{e}/2$ and the number of monomers is $N=N_{b}+N_{p}$, one can
see that Eq.$\left( \ref{zzh}\right) $ is identical to Eq.$\left( \ref{Z}%
\right) $ if the following identification is made 
\begin{eqnarray*}
J &=&e^{\mu } \\
h &=&e^{-\epsilon _{e}}J^{1/2}\equiv h_{0}J^{1/2}
\end{eqnarray*}
Each term in the sum in Eq.(\ref{zzh}) corresponds to a different
realization of the grand canonical ensemble of a solution of self-avoiding
chains. We have thus demonstrated that the grand canonical partition
function of a solution of polydisperse, living polymers is identical to that
of the $n$-component Heisenberg model where the number of components $%
n\rightarrow 0$. As mentioned in the Introduction, the lattice approximation
is adequate as long as the chains are much longer than the lattice constant.
This theory describes the\textit{\ }ensemble of\textit{\ self-assembling}
`living' chains, which we are of interest to us. The molecular weight and
the length distribution are not fixed but are a functions of external
parameters, such as the monomer density. However, this model has also been
successfully applied to non-self-assembling polymer solutions (where the
degree of polymerization is constant and fixed by the chemical preparation
technique). The continuum version of this model is the basis for the
application of field-theoretical methods to polymer physics \cite
{descloiseaux,witten,vilgis}. The self-assembling theory can be used to
predict the properties of non-self-assembling polymer solutions of long
chains (small $h$ in our formulation), because the scaling behavior of such
systems is universal and independent of the detailed chain properties.

\subsection{Three-fold junctions\label{juncn0}}

The model presented in the previous section can be modified to include the
possibility of junctions that connect several chains; this allows a
formation of a branched structure\ and is achieved by introducing an
additional term to the Hamiltonian of Eq.$\left( \ref{H}\right) $. Again, we
consider the grand canonical partition function of an equilibrium solution
of branched cross-linked polymers with reversible cross-links. The
cross-links (branching points) are in thermal equilibrium and can break and
reform. Similar to the solution of polymer chains discussed above, the grand
canonical partition function of the system is

\begin{equation}
Z_{3}=\sum_{\{N,N_{e},N_{j}\}}\exp [\mu N/T]\,\exp \left[ -\epsilon
_{e}N_{e}/T\right] \,\exp [-\epsilon _{j}N_{j}/T]\,\mathcal{N}(N,N_{e},N_{j})
\label{z3}
\end{equation}
where $\epsilon _{j}$ is the energy of a cross-link relative to the energy
of the monomer in the middle of a chain, $N_{j}$ is the number of junctions
(cross-links) and $N(N,N_{e},N_{j})$ is the number of ways to put
self-avoiding branched chains of total length $N$ on a lattice so that $%
N_{e} $ ends and $N_{j}$ junctions are formed. The chemical potential for
the monomers, $\mu $ and the end energy $\epsilon _{e}$ are defined as in
the previous section. Similar to the discussion of the end energy and its
associated chemical potential, the junction energy taken with the opposite
sign, -$\epsilon _{j},$ can be interpreted as the chemical potential
conjugate to the number of junctions. We suppose that the physically
controllable parameters are the $monomer$ density $\phi $, the temperature $%
T $ and the ends/junctions energies $\epsilon _{e},\epsilon _{j}$. In this
formulation, the density of ends $\phi _{e}$ and the density of junctions $%
\phi _{j}$ are not fixed but determined in thermal equilibrium as functions
of $\phi $ and $T$.

In this section we show how the three-fold junctions can be described by `$%
n=0$' model. The generalization to junctions of arbitrary functionality is
discussed in the next section. As in the previous section, one can relate
the partition function of Eq.$\left( \ref{z3}\right) $ to an equivalent spin
model on a lattice. The mapping accounts rigorously for all possible
configurations of branched self-avoiding clusters, except for closed rings.
The influence of rings on the properties of living polymers has been studied
in Ref. \cite{petschek}. However, in the case of branched chains, the
influence of closed rings on thermodynamic properties of the system is
small, except at very low densities, and they are not treated in this paper.
The reason for this is that for any closed ring there is an exponentially
big number of branched clusters formed by attaching side chains to it. No
topological information regarding, e.g., the network formation, or the
cluster size distribution can be extracted from the model formulated below.
However, it can be modified to include the treatment of topological
properties too, as shown in Sec.\ref{spinspin}.

We introduce the following Hamiltonian that contains a term which couples
three adjacent spins:

\begin{equation}
H_{3}=\sum_{i}\vec{h}\cdot \vec{S}_{i}+\sum_{ij}J(\vec{S}_{i}\cdot \vec{S}%
_{j})+\sum_{ijk}KS_{1,i}S_{1,j}S_{1,k}  \label{h3}
\end{equation}
where $i,j,k$ sum is over all distinct triplets on the lattice. The field $h$
is chosen to point in the direction $(1,0...0)$. The partition function
corresponding to this Hamiltonian is

\begin{equation}
Z_{H_{3}}=\text{Tr}_{\left\{ S_{i}\right\} }\exp [-H_{3}\{S_{i}\}/T^{\star }]
\label{trh3}
\end{equation}
The `temperature' $T^{\star }$ can be taken equal to unity and any other
choice would amount to a redefinition of the constants $J,K$ and $h.$
Proceeding as in the previous section, we expand $Z_{H_{3}}$ in series in
powers of $K,J$ and $h$. In the limit $n\rightarrow 0$ the terms in the
expansion of $Z_{H_{3}}$ in powers of $J,K$ and $h$ are similar to those
discussed in the previous section. However, there are additional terms
obtained by inserting into any term of the expansion of Eq.$\left( \ref
{expansionterms}\right) $ combinations of the form

\begin{equation}
...JS_{j^{\prime },1}S_{j,1}\text{{\Large [}}\underset{\text{junction}}{%
\underbrace{KS_{j,1}S_{m,1}S_{n,1}}\text{{\large ]}}}JS_{n,1}S_{n^{\prime
},1}JS_{m,1}S_{m^{\prime },1}...  \label{tree}
\end{equation}
Each insertion of this kind corresponds to a three-fold junction on a
lattice, joining the sites $j,m,n$ (cf. Fig.\emph{\ }\ref{juncfig}). Thus, a
general term in the expansion of $Z_{H_{3}}$ might look like 
\begin{equation}
\text{Tr }hS_{i,1}JS_{i,1}S_{j,1}\underset{\text{junction}}{\underbrace{%
[KS_{j,1}S_{n,1}S_{k,1}]}}JS_{n,1}S_{n^{\prime },1}hS_{n^{\prime
},1}...JS_{k,1}S_{m,1}hS_{m,1}  \label{juncexpansion}
\end{equation}
This particular term, for example, corresponds to a three-fold branched,
self-avoiding chain, which has a junction between the points $j,n,k$, and
free ends at points $i,n^{\prime }$ and $m$. In other words, the partition
function is

\begin{equation}
Z_{H_{3}}=\text{Tr }\exp
[-H_{3}]=\sum_{\{N_{b,}N_{e},N_{j}\}}J^{N_{b}}h^{N_{e}}K^{N_{j}}\text{ }%
\mathcal{N}(N_{b},N_{e},N_{j})  \label{zh3}
\end{equation}
where $\mathcal{N}(N_{b},N_{e},N_{j})$ is the number of ways to arrange on a
lattice an ensemble of\ self-avoiding branched random walks with a total
number of $bonds$ $N_{b}$ (excluding three `ghost' bonds at each junctions,
cf. Fig. \ref{juncfig}), a total number of ends $N_{e},$ and a total number
of junctions $N_{j}$. The closed rings fall out form the expansion \ref{zh3}
due to summation over \ spin components. Each ring produces a contribution
proportional to $\ n$ which tends to zero as $\ n\rightarrow 0$, analogously
to the case of linear chains. From simple geometric considerations, the
number of monomers, $N$, is always $N=N_{b}+\frac{1}{2}N_{e}+\frac{3}{2}%
N_{j} $ (we do not count the junctions as monomers; see Appendix \ref{alpha}%
), regardless of the presence or absence of loops. Note that this
formulation does not treat the junctions as point-like objects. Choosing
different forms of three-spin coupling in Eq.$\left( \ref{h3}\right) $ one
can study the influence of the junction size on the system properties (cf.
Appendix \ref{alpha}). If the following identifications are made

\begin{eqnarray}
J &=&e^{\mu /T}  \label{corresp} \\
h &=&e^{-\epsilon _{e}/T}J^{\frac{1}{2}}\equiv h_{0}J^{\frac{1}{2}}  \notag
\\
K &=&e^{-\epsilon _{j}/T}J^{\frac{3}{2}}\equiv K_{0}{}J^{\frac{3}{2}}  \notag
\end{eqnarray}
then the $Z_{H_{3}}$ of Eq.$\left( \ref{zh3}\right) $ becomes identical to $%
\ Z_{3}$ of Eq.$\left( \ref{z3}\right) $. To summarize, we have shown that
in the $n\rightarrow 0$ limit, the grand-canonical ensemble of a solution of
branched, reversibly cross-linked polymers is equivalent to a Heisenberg
magnet with three-spin term. It is important to emphasize that although the
closed rings ( linear, unbranched, chains closed to form a ring) are not
included in the expansion (\ref{juncexpansion}), all the intracluster loops
are counted properly.

The concentrations of the monomers, ends and junctions, $\phi ,$ $\phi _{e}$
and $\phi _{j}$ respectively can be obtained by differentiating the
partition function with respect to a relevant parameters as follows from
Eqs.(\ref{z3},$\ref{zh3}$):

\begin{eqnarray}
\phi &=&\frac{1}{V}\,\frac{\partial \ln Z}{\partial \ln J};\text{ \ }\phi _{%
\text{bonds}}=\frac{1}{V}\frac{\partial _{s}\ln Z}{\partial _{s}\ln J}
\label{densities} \\
\text{ }\phi _{e} &=&\frac{1}{V}\frac{\partial \ln Z}{\partial \ln h}\text{; 
}\phi _{j}=\frac{1}{V}\frac{\partial \ln Z}{\partial \ln K}  \notag
\end{eqnarray}
\emph{\ \ }where $\partial _{s}$\ denotes a derivative with respect to $J$
with the generalized fugacities $K,J$ and $h$ taken to be independent. The
great advantage of the present formulation is that it allows to apply the
methods of statistical mechanics developed in the past four decades for the
treatment of the spin models.

In some systems, junctions may form spontaneously; in that case, the number
of junctions, distribution of the chain intervals between the junctions and
the branched clusters' size will be determined by the junction energy $%
\epsilon _{j},$ the monomer density $\phi $ and the temperature $T$. This is
what one expects, for example, in dipolar fluids or microemulsions. In other
systems, junction formation may only be possible if linker molecules, that
connect several chains, are added to the system. In this case, the grand
canonical formulation presented here amounts to the assumption that the
system is in equilibrium with a bath of linker molecules with chemical
potential -$\epsilon _{j}$.

The grand canonical potential per unit volume, $\omega (\epsilon
_{j},\epsilon _{e},\mu ),$ is given by 
\begin{equation}
\omega (\epsilon _{j},\epsilon _{e},\mu )/T=-\frac{1}{V}\ln Z(K_{0},h_{0},J)
\label{Omega1}
\end{equation}
Other physically realizable situations are related to Eq.($\ref{Omega1})$ by
a Legendre transform. For instance, the free energy corresponding to the
case where junctions form only in the presence of linker molecules is 
\begin{equation*}
g(\mu ,\epsilon _{e},N_{j})=\omega -\phi _{j}\ln K_{0}
\end{equation*}
Of course, in the thermodynamic limit ($V\rightarrow \infty )$ the
properties of the system are identical in either ensemble. Here, we will
focus on the case where the physically controllable parameters are the total
monomer density, $\phi ,$ and the defect energies, $\epsilon _{e}/T=-\ln
(h_{0})$ and $\epsilon _{j}/T=-\ln (K_{0})$. The corresponding Helmholtz
free energy per unit volume is 
\begin{equation*}
f=\omega +\mu \phi =\omega +\phi \ln J
\end{equation*}
\FRAME{ftbpFU}{3.365in}{2.559in}{0pt}{\Qcb{Junction can be formed between
three adjacent sites $\ i,j,k$. The Boltzmann factor for a junction is $%
K_{0}=e^{-\protect\epsilon _{j}/T}$.}}{\Qlb{juncfig}}{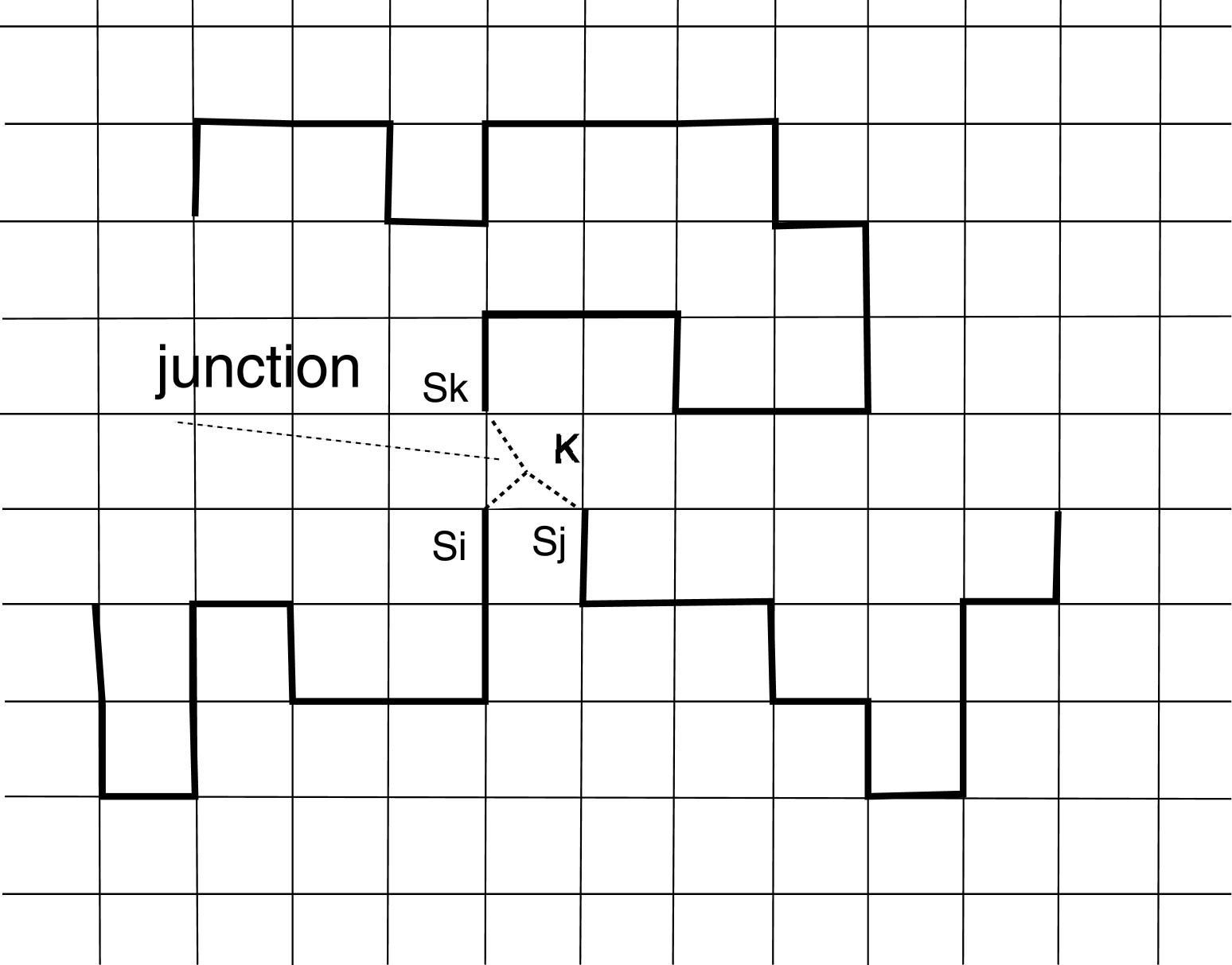}{\special%
{language "Scientific Word";type "GRAPHIC";display "USEDEF";valid_file
"F";width 3.365in;height 2.559in;depth 0pt;original-width
32.2497in;original-height 25.2707in;cropleft "0";croptop "1";cropright
"1";cropbottom "0";filename 'L:/networks/junclat.jpg';file-properties
"NPEU";}}

\section{Mean field theory\label{mf}}

\subsection{Mean field approximation}

The first approximation in evaluating the free energy is to disregard the
spatial variations of the densities and the long range correlations between
the monomers. In this section we assume that the system is spatially uniform
and calculate the free energy as a function of the average monomer volume
fraction $\phi $, and the ends and junction densities, $\phi _{e}$ and $\phi
_{j}$, respectively. A mean-field calculation of the free energy for a
system with no junctions has been presented in Refs. \cite
{wheeler,wheelersulf}. We extend this calculation by an alternative method
to a system that contains both ends and junctions. We first calculate the
grand canonical potential $\Omega /T=-\ln Z$ in the mean field
approximation, which is generally known to predict the qualitative
thermodynamic behavior correctly\cite{huang}. As we shall show subsequently
in Sec. \ref{spinspin}, the mean field approximation neglects the internal
loops in the branched clusters, because only local properties are preserved
in the mean field approximation. Geometrically, it is equivalent to Flory
construction \cite{florybook} when each cluster is constructed by adding
consecutive bonds, disregarding the positions of the previously placed ones,
and the long range correlations are lost. In particular, intra-cluster
self-avoidance is neglected in the mean field approximation. However, the
excluded volume between different clusters is taken into account. The
partition function $Z$ is given by 
\begin{equation}
Z=\text{Tr }\exp \text{{\Large [}}-\sum_{i,j}J\vec{S}_{i}\cdot \vec{S}%
_{j}-\sum_{i,j,k}KS_{1,i}S_{1,j}S_{1,k}-\sum_{i}\vec{h}\cdot \vec{S}\text{%
{\Large ]}}  \label{mf1}
\end{equation}
The sums in the argument of the exponent of Eq.$\left( \ref{mf1}\right) $
are over $distinct$ pairs and triples, as explained in Sec. \ref{juncn0}. We
now note the following identity 
\begin{equation*}
\vec{S}_{i}\vec{S}_{j}=\underline{(\vec{S}_{i}-\vec{S})(\vec{S}_{j}-\vec{S})}%
-\vec{S}^{2}+\vec{S}(\vec{S}_{i}+\vec{S}_{j})\simeq -\vec{S}^{2}+\vec{S}(%
\vec{S}_{i}+\vec{S}_{j})
\end{equation*}
where $\vec{S}=\langle \vec{S}_{i}\rangle .$ The underlined term in the
equation is quadratic in the deviation of the local spin from its average
value $S$. It is this term that is neglected in the mean field
approximation. As the 'magnetic field' $h$ has been chosen to point in the
direction $\left( 1,0..0\right) $ the only non-zero component of the average
spin is $S_{i,1}$. Therefore, the transverse components of the spin $\vec{S}$
do not contribute to the mean field approximation, due to the fact that $%
\langle S_{i,\bot }\rangle =0$. So, in the following we drop the vector sign
and $S_{i}$ signifies the component of $\vec{S}_{i}$ in the direction of $h$%
. \ Similarly, the mean field approximation for the three-spin term is:

\begin{equation*}
S_{i}S_{j}S_{k}\simeq -2S^{3}+S(S_{i}+S_{j}+S_{k})
\end{equation*}
Thus, the mean field approximation to the Hamiltonian reads 
\begin{equation*}
H_{\text{MF}}=V(\frac{1}{2}qJS^{2}+2\alpha KS^{3})-\sum_{i}(qJS+3\alpha
KS^{2}+h)S_{i}
\end{equation*}
where $V$ is the number of lattice sites (which tends to infinity in the
thermodynamic limit) and the sum is over all lattice sites; $q$ is the
lattice coordination number. The prefactor $\alpha =\frac{1}{3}\alpha
^{\prime }$, where $\alpha ^{\prime }$ is the number of possible
configurations of $j,k$ \ in the triple term $i,j,k$ making a junction
around the cite $i$. The exact value of $\alpha ^{\prime }$ depends on the
type of the lattice used and on the physical assumptions made in counting
the physically relevant junction configurations. Physically, it represents
the entropy associated with local rearrangements of a single junction and we
discuss its value for semi-flexible or rigid systems in Sec.\ref{rigid}. In
most of this paper, we shall use the\ value of $\alpha ^{\prime }$
appropriate for a simple cubic lattice $\alpha ^{\prime }=q(q-2)/3$ (cf.
Appendix \ref{alpha}). A particular feature of the $n\rightarrow 0$ limit,
following from Eq.(\ref{n0}) \cite{degenbook,sarma,wheeler}(cf. also
Appendix \ref{n=0app}) is the fact that for an arbitrary vector $\vec{k},$%
\begin{equation*}
\lim_{n\rightarrow 0}\text{Tr }e^{\vec{k}\vec{S}_{i}}=1+\frac{1}{2}k^{2}
\end{equation*}
Consequently, performing the trace in Eq.(\ref{mf1}), we get 
\begin{equation}
\frac{1}{V}(\Omega /T)\equiv \omega =\frac{1}{2}qJS^{2}+2\alpha KS^{3}-\ln 
{\LARGE (}1+\frac{1}{2}(qJS+3\alpha KS^{2}+h)^{2}{\LARGE )}  \label{Omeg}
\end{equation}
The average value of the spin, $S,$ is given by 
\begin{equation*}
S=-\frac{\partial \omega }{\partial h}=\frac{1}{V}\langle
\sum_{i}S_{i}\rangle
\end{equation*}
The total monomer density is 
\begin{equation}
\phi =\phi _{bonds}+\frac{1}{2}\phi _{e}+\frac{3}{2}\phi _{j}=-\frac{%
\partial _{s}\omega }{\partial _{s}\ln J}-\frac{1}{2}\frac{\partial \omega }{%
\partial \ln h}-\frac{3}{2}\frac{\partial \omega }{\partial \ln K}=-\frac{%
\partial \omega }{\partial \ln J}  \label{phi}
\end{equation}
as follows from Eq.(\ref{densities}), and can be also understood by a simple
geometrical argument (cf. Appendix \ref{alpha}). Finally, combining all the
equations we find:

\begin{eqnarray*}
\omega &=&\frac{1}{2}qJS^{2}+2\alpha KS^{3}-\ln {\LARGE (}1+\frac{1}{2}%
(qJS+3\alpha KS^{2}+h)^{2}{\LARGE )} \\
S &=&-\frac{\partial \omega }{\partial h}=\frac{qJS+3\alpha KS^{2}+h}{1+%
\frac{1}{2}(qJS+3\alpha KS^{2}+h)^{2}} \\
\phi &=&\frac{1}{2}S\underset{x}{(\underbrace{qJS+3\alpha KS^{2}+h}})
\end{eqnarray*}
Denoting $qJS+3\alpha KS^{2}+h$ by $x,$ the solution of this system of
equations gives 
\begin{eqnarray*}
\phi &=&\frac{1}{2}Sx \\
S &=&\frac{x}{1+\phi x/S}
\end{eqnarray*}
Consequently 
\begin{eqnarray*}
x &=&\sqrt{\frac{2\phi }{1-\phi }};\text{ \ }1+\frac{1}{2}x^{2}=\frac{1}{%
1-\phi };\text{ }S=\sqrt{2\phi (1-\phi )} \\
\omega &=&qJ\phi (1-\phi )+2\alpha K(2\phi )^{3/2}(1-\phi )^{3/2}+\ln
(1-\phi )
\end{eqnarray*}
It follows that in the limit of systems with a relatively small number of
junctions and ends, compared to the total number of monomers in the chains, (%
$h_{0}(2\phi )^{1/2}\ll \phi $, $K_{0}(2\phi )^{3/2}\ll \phi $),

\begin{equation}
qJ=(x-h-3\alpha KS^{2})/S\simeq \frac{1}{1-\phi }{\LARGE (}1-\frac{h_{0}}{%
\sqrt{2\phi }}-3\alpha K_{0}\sqrt{2\phi }{\LARGE )}  \label{qJ}
\end{equation}
Because we are interested in those systems where the density, $\phi ,$ (and
not the Lagrange multiplier, $J,)$ is a physically controllable parameter,
we perform a Legendre transform $f=\omega +\phi \ln qJ$ in order to obtain
the Helmholtz free energy per unit volume $f(\phi ,h_{0},K_{0}).$ Up to the
first order in $h_{0}$ and $K_{0},$we find:

\begin{equation}
f/T\simeq (1-\phi )(\ln (1-\phi )-1)-\frac{h_{0}(2\phi )^{1/2}}{q^{1/2}}-%
\frac{\alpha K_{0}(2\phi )^{3/2}}{q^{3/2}}  \label{FF}
\end{equation}
where $q$ is the coordination number of the lattice. The first term in Eq.$%
\left( \ref{FF}\right) $ describes the self-avoidance of the chains. The
second and the third term are the densities of ends and junctions,
respectively, as can be seen from Eqs.(\ref{densities}). Denoting the volume
fraction of ends by $\phi _{e}$ and of junctions by $\phi _{j}$ it follows: 
\begin{eqnarray}
\frac{h_{0}(2\phi )^{1/2}}{q^{1/2}} &=&\phi _{e}  \label{defsoffi} \\
\frac{\alpha K_{0}(2\phi )^{3/2}}{q^{3/2}} &=&\phi _{j}  \notag
\label{juncoffi}
\end{eqnarray}
\qquad The expansion in powers of $h_{0}$ and $K_{0}$ is justified because
we are interested in the limit where the density of ends is much smaller
than the total monomer density, $\phi _{e}\ll \phi $. The density of
junctions is always smaller than the total monomer density, $\phi _{j}\ll
\phi $ as follows form Eq.$\left( \ref{defsoffi}\right) $ due to the fact
that $K_{0}=e^{-\epsilon _{j}/T}<1$ for $\epsilon _{j}>0$.

We note in passing that the choice of the lattice affects only the numerical
prefactors in Eqs.($\ref{defsoffi}$),(\ref{FF}). The lattice-specific
dependence of junctions and ends densities in Eq.(\ref{defsoffi}) on $q$ is
consistent with a simple probabilistic argument. In equilibrium the
probabilities of defect (e.g., ends and junctions) formation and break-up
must be equal. The probability of bond formation is proportional to $\frac{1%
}{2}\phi _{e}^{2}q,$ because a bond can be formed whenever two ends are
neighbors on a lattice (the factor $\frac{1}{2}$ is due to
indistinguishability of any two ends).\ The probability of bond break-up is
proportional to the total number of bonds $N_{b}\approx N$. Taking into
account that the formation of two ends from a single bond costs an energy $%
2\epsilon _{e},$ the probability of a bond break-up is $\phi $ $%
e^{-2\epsilon _{e}/T}.$ Equating the probabilities of bond breaking and bond
formation produces Eq.(\ref{defsoffi}) (recalling that $h_{0}=e^{-\epsilon
_{e}/T})$. Similarly, three ends can coalesce to form a junction, giving in
equilibrium $\phi _{e}^{3}/e^{-3\epsilon _{e}}\simeq \alpha \phi
_{j}/e^{-\epsilon _{j}}$ ( $K_{0}=e^{\epsilon _{j}/T}).$ Note that this
argument is independent of the fact whether the 'ground state' consists of
infinite chains or closed rings of linear chains.

Together with Eq.(\ref{defsoffi}), the free energy $f$ of Eq.(\ref{FF}) can
be cast into the lattice-independent form 
\begin{equation}
f/T=\phi +(1-\phi )\ln (1-\phi )-\phi _{e}-\phi _{j}  \label{Fdef}
\end{equation}
The first term in Eq.(\ref{Fdef}) is due to self-avoidance between the
chains. This particular form of $f$ is expected on very general grounds. If
one thinks of the ends and junctions as defects in the system of infinite
chains, each defect lowers the free energy by $k_{B}T$, which is true for
any system with non-conserved defects, e.g. dislocations in crystal
structures.

\subsubsection{Average distance between the defects\label{lbarsec}}

The average volume fractions of junctions, $\phi _{j},$ and ends, $\phi
_{e}, $ are not fixed but depend on the total volume fraction of monomers, $%
\phi $. Similarly, the self-assembling nature of the system means that the
size distribution of the branched aggregates is polydisperse.

Absorbing the lattice-dependent prefactors in Eq.$\left( \ref{defsoffi}%
\right) $ into the definition of the constants $K$ and $h,$ it is rewritten
as: 
\begin{equation}
\phi _{j}=K_{0}^{\prime }\phi ^{3/2}\text{ \ and }\phi _{e}=h_{0}^{\prime
}\phi ^{1/2}  \label{def1}
\end{equation}
As can be seen from Eq.$\left( \ref{def1}\right) $ for $\phi \ll
h_{0}^{\prime }/K_{0}^{\prime },$the number of junctions is much smaller
than the number of ends, $\phi _{j}\ll \phi _{e}$ while in the opposite
limit $\phi _{j}\gg \phi _{e}.$ Let us consider the mean length of a chain
segment in between two consecutive defects (i.e., ends or junctions), $\bar{l%
}$. Each junction is attached to three chain segments while each end is
attached to a single segment. The total length of the segments is
proportional to the total volume of monomers in the system. Consequently ($%
\frac{3}{2}N_{j}+\frac{1}{2}N_{e})\bar{l}=N$; the factor $\frac{1}{2}$
corrects for the double counting of each segment. Consequently, we find: 
\begin{equation}
\bar{l}=\frac{2\phi }{3\phi _{j}+\phi _{e}}\simeq \frac{2\phi }{%
3K_{0}^{\prime }(2\phi )^{3/2}+h_{0}^{\prime }(2\phi )^{1/2}}  \label{lbar}
\end{equation}
For $\phi \ll h_{0}^{\prime }/K_{0}^{\prime }$, the ends dominate and the
mean chain length in between end points is $\bar{l}\sim \phi ^{1/2}$ while
for $\phi \gg h_{0}^{\prime }/K_{0}^{\prime }$ the junctions dominate and
the mean chain length in between junction points is $\bar{l}\sim \phi
^{-1/2}.$ These results are in agreement with the known results for both
non-branched micelles and pure networks \cite{sambook,drye}. These results
suggest, that at high densities (where the ends are negligible), a connected
network is formed, as we shall actually prove in Sec. \ref{transversal}.
They also imply that the mean chain length between defects, $\bar{l}(\phi )$
has a maximum around $\phi \sim h_{0}^{\prime }/K_{0}^{\prime }$. One can
easily convince oneself taking the derivative $\frac{\partial \bar{l}}{%
\partial \phi }$ that $\bar{l}$ is indeed a non-monotonic function of $\phi $%
\ whose maximum is located at $\phi =\frac{1}{2}\frac{h_{0}^{\prime }}{%
3K_{0}^{\prime }}$ or, in other words, where $\phi _{e}=3\phi _{j}$. This
fact will become important for the determination of the details of the
evolution of the system from a state of disconnected chains to that of a
connected network (cf. Sec. \ref{transversal}).

In the absence of junctions (e.g. when the junction energy is very high and
the resulting value of $K$ is close to zero), the mean chain length in
between end points is $\bar{N}=\frac{\phi }{2\phi _{e}}=h_{0}^{\prime }\phi
^{1/2}$. The free energy $F(\phi ,h_{0})$ of Eq.(\ref{Fdef}) can be
transformed to a free energy that depends only on $\phi $ and $\phi _{e}$
(or the mean chain length $\bar{N})$ by means of the Legendre transformation 
$f(\phi ,h_{0})+\phi _{e}\ln h_{0}=\bar{f}(\phi ,\bar{N}).$ The result is
identical to the Flory-Huggins expression for the free energy of the polymer
solutions: 
\begin{equation*}
\bar{f}(\phi ,\bar{N})=(1-\phi )\ln (1-\phi )+\frac{\phi }{\bar{N}}\ln \phi
\end{equation*}
As mentioned above, this is because the mean field approximation is
geometrically equivalent to Flory lattice construction, which disregards the
intra-chain self-avoidance, but retains the excluded volume interactions
between different chains. Thus our model describes the case of polydisperse
chains with both ends and junctions as well as the limiting case of finite
chains (with no junction points) where the average length is well defined;
this is the case that is applicable to chemically prepared polymeric chains.

\subsection{\protect\bigskip The junctions-ends transition\label{j-e}}

The free energy $f$ of Eq.(\ref{FF}) is unstable in a certain range of
parameters ( monomer density $\phi ,$ defect energies $\epsilon _{e}$ and $%
\epsilon _{j}$ and temperature) and shows a first-order phase transition,
terminating at a critical point. The spinodal line associated with this
transition is determined by the condition $\frac{\partial ^{2}F}{\partial
\phi ^{2}}=0\ $and at the critical point, $\frac{\partial ^{3}F}{\partial
\phi ^{3}}=0$. We determine the conditions for the critical point from our
model as: 
\begin{eqnarray}
\frac{\partial ^{2}F}{\partial \phi ^{2}} &=&\frac{1}{1-\phi }+\frac{h_{0}}{%
q^{1/2}}(2\phi )^{-3/2}-3\alpha \frac{K_{0}}{q^{3/2}}(2\phi )^{-1/2}=0
\label{secder} \\
\frac{\partial ^{3}F}{\partial \phi ^{3}} &=&\frac{1}{(1-\phi )^{2}}-3\frac{%
h_{0}}{q^{1/2}}(2\phi )^{-3/2}+3\alpha \frac{K_{0}}{q^{3/2}}(2\phi )^{-3/2}=0
\notag
\end{eqnarray}
Recalling that $h_{0}=e^{-\epsilon _{e}/T}$ and $K_{0}=e^{-\epsilon _{j}/T}$%
, the Eq.(\ref{secder}) can be solved analytically for small $\phi $,
expanding $\left( 1-\phi \right) (\ln \left( 1-\phi \right) -1)\simeq \frac{1%
}{2}\phi ^{2},$ which gives 
\begin{eqnarray}
T_{c} &=&\frac{-3\epsilon _{j}+\epsilon _{e}}{\ln q^{4}/(4\alpha ^{3})}
\label{crit} \\
\phi _{c} &=&\frac{1}{2}\frac{q}{\alpha }e^{(\epsilon _{j}-\epsilon
_{e})/T_{c}}  \notag
\end{eqnarray}
while for arbitrary $\phi $ they can be solved numerically (Fig.\ref{corrfig}%
). It is interesting to note that in the small density limit, $\phi
_{e}\approx \phi _{j}$ at the critical point, which emphasizes the fact that
the transition is junctions-induced. For $T>T_{c}$ the system is a \emph{%
homogeneous} mixture of chains and branched aggregates, while for $T<T_{c}$
there is a two-phase equilibria between an end-rich phase that coexists with
a junction-rich phase. As will be shown in Sec. \ref{transversal}, the
end-rich phase usually consists of dilute, disconnected chains, while the
junction-rich phase is usually a \textit{connected network} that spans the
whole system volume.

Although there are \textit{no direct\ interactions} between the monomers in
our model, junction formation induces effective \textit{attraction} between
the monomers, and it is this effective interaction that leads to the phase
separation of Eq.$\left( \ref{secder}\right) .$ The Eq.$\left( \ref{crit}%
\right) $ has several interesting consequences: First, the phase separation
is possible only when $\epsilon _{j}<\frac{1}{3}\epsilon _{e}$ (as long as $%
q^{4}/(4\alpha ^{3})>1);$ in the opposite case, the junctions are present
only as a minor `perturbation' to a system of linear chains and their number
is too small to generate an attraction large enough to drive a macroscopic
phase separation. Second, the critical monomer density $\phi _{c}<\frac{1}{2}
$; this fact is important for the understanding of the correlations between
monomers and between junctions in the system. Third, if $\alpha $ is large ($%
q^{4}/(4\alpha ^{3})<1)$, there is no upper critical temperature, and the
phase separation takes place even at infinitely high temperatures ($%
T_{c}\rightarrow \infty )$. If the `microscopic' end cap and junction
energies, $\epsilon _{e}$ and $\epsilon _{j}$ themselves depend on
temperature or density (e.g., in microemulsions \cite{tsvimicro}), the phase
diagram may become more complicated. Note that the phase transition
described by Eq.(\ref{secder}) shows reentrant behavior. At high
temperatures, there is no phase separation, as usual. At very low
temperatures the parameter $K_{0}=e^{-\epsilon _{j}/T}$ tends to zero and
there is no separation \textit{either}, as shown in Fig. \ref{fig1}, because
the number of thermally generated junctions is too small to drive a phase
separation.

The phase transition discussed here is of purely entropic origin, which
augments a qualitative claim by De Gennes that the cross-links are
thermodynamically equivalent to attractive interactions \cite{degenbook}.
The origin of this phase transition lies in the fact that although the
translational entropy of the chains is lower in the dense phase, the total
entropy is still $\ higher$ due to high entropy of the self-assembled
junctions, which are abundant in the dense phase. It is important to
emphasize that the presence of junctions induces an attraction between all 
\textit{monomers} and not \emph{only }between the junctions, as we shall
show in Sec. \ref{spinspin}. Also, the phase separation discussed here is
not the same as the percolation (connectivity) transition, where an infinite
cluster appears, although at both transitions a macroscopic connected
network is formed. As we shall see in Sec. \ref{transversal}, the
percolation transition, at which an infinite branched cluster appears in the
system, is not a thermodynamic, but rather\emph{\ }a \textit{structural
transition}. In some cases, the percolation transition can be masked by the
end-junction phase separation. It is also interesting to note that for the
values of $\epsilon _{e}$ and $\epsilon _{j}$ at which the phase separation
transition is absent ($h_{0}>K_{0})$, there is also no maximum in $\bar{l}%
(\phi )$, as follows from Eq.$\left( \ref{lbar}\right) $.

\subsubsection{Osmotic pressure and the phase coexistence line}

The osmotic pressure $\Pi $ can be calculated from the free energy $f$ of Eq.%
$\left( \ref{FF}\right) $:

\begin{equation*}
\Pi =-f+\frac{\partial f}{\partial \phi }\phi \simeq -\ln (1-\phi )-\phi +%
\frac{1}{2}h_{0}^{\prime }(2\phi )^{1/2}-\frac{1}{2}K_{0}^{\prime }(2\phi
)^{3/2}
\end{equation*}
Note that the coefficient before $h_{0}^{\prime }$-dependent (ends) term is 
\textit{positive} due to the fact that the exponent in $\phi ^{1/2}$ is less
than one, while the sign of the junctions term is negative. This emphasizes
that fact that, although both ends and junctions terms are negative in the 
\textit{free energy}, the junctions term is thermodynamically equivalent to
an\textit{\ attraction,} while the ends term produces an effective \textit{%
repulsion.} The location of the phase coexistence line is determined as
usual requiring that the chemical potentials $\mu =\frac{\partial f}{%
\partial \phi }$ and the osmotic pressures $\Pi $ in both phases are equal: 
\begin{equation*}
\mu (\phi _{1})=\mu \left( \phi _{2}\right) \text{ \ }\Pi \left( \phi
_{1}\right) =\Pi \left( \phi _{2}\right)
\end{equation*}
Numerical solution of these equations is shown in Fig. \ref{fig1}.

\subsection{Four-fold and higher functionality junctions: gelation vs.
vulcanization\label{vulc}}

The analysis of the previous section can be generalized to systems with
cross-links of arbitrary functionality $z,$ by the addition to the
Hamiltonian H$_{3}$ of Eq.(\ref{h3}) a term that couples $z$ spins. 
\begin{equation*}
-\sum_{i_{z}}K_{z}\underset{z\text{ times}}{\underbrace{%
S_{1,i_{1}}S_{1,i_{2}}..S_{1,i_{z}}}}
\end{equation*}
with $K_{z}=J^{z/2}e^{-\epsilon _{z}}$ (cf. Sec. ; $\epsilon _{z}$ being the
energy of the $z$-fold junction. In this case the resulting Helmholtz free
energy per unit volume then takes the form 
\begin{equation}
f/T\simeq (1-\phi )(\ln (1-\phi )-1)-\alpha _{1}h_{0}\phi ^{1/2}-\alpha
_{z}e^{-\epsilon _{z}}\phi ^{z/2}  \label{zfold}
\end{equation}
where $\alpha _{1}$ and $\alpha _{z}$ are two numerical, lattice dependent
prefactors reflecting the local ends and junctions \ configurations,
respectively. There are two interesting points to be noted.

$\left( i\right) $ There is no thermodynamic phase separation for $z>4$, as
can be readily seen from Eqs.$\left( \ref{secder},\ref{zfold}\right) ,$
because the resulting free energy of Eq. (\ref{zfold}) is always convex ($%
\frac{\partial ^{2}f}{\partial \phi ^{2}}>0)$. The attraction induced by the
junctions weakens with increasing $z.$ The topological percolation
transition is, however, still present, as discussed in Sec. \ref{transversal}

$\left( ii\right) $ When $z=4$, the four-fold junctions contribution to the
free energy is indistinguishable from the two-body attractions between the
monomers (at least at the mean-field level), both contributing term $\sim
\phi ^{2}$ to the free energy for $\phi \ll 1$. The free energy is 
\begin{equation}
f/T\simeq (\frac{1}{2}-\frac{1}{2}e^{-\epsilon _{4}^{\prime }})\phi
^{2}-\phi _{e}  \label{4fold}
\end{equation}
where the effective \ junction energy $\epsilon _{4}^{\prime }$ has been
redefined to absorb the internal degrees of freedom of a junction, expressed
in the prefactor $\alpha $ of Eq.(\ref{zfold}).The term $\frac{1}{2}%
e^{-\epsilon _{4}^{\prime }}$ corresponds to the $\chi $ parameter of the
Flory theory. At the point where $\epsilon _{4}^{\prime }=0,$ the excluded
volume repulsion is balanced by the junction-induced attraction, a situation
analogous to the $\Theta $-point of polymer solutions.\ The location of the
compensation point is intuitively very clear: if $\epsilon _{4}^{\prime }=0,$
there is no cost for creating a junction and the chains can intersect each
other at will (the fact that the junctions are four-fold is crucial!),
thereby acting as completely Gaussian phantom chains.

The three- and four-fold junctions correspond to two physically distinct
situations. Three- fold junctions describe the process of gelation when the
monomers simultaneously cross-link and polymerize in the reaction bath. The
clusters formed this way exhibit mostly three-fold junctions, simply due to
the fact that the collision of three monomers is more probable that the
four-fold collision. In addition, in some systems (e.g. microemulsions \cite
{tsvimicro}) it can be shown that the energy of four-fold junctions is
higher than that of three-fold junctions. The four-fold cross-links
correspond to the process of vulcanization, when existing long polymer
chains are cross-linked by addition of cross-link molecules, by irradiation
or other means \cite{isaacson,doibook,florybook}. Formation of three fold
junctions is structurally inhibited by the nature of the cross-link which
cross-links two preexisting chains. Contrary to gelation, these systems show
only four-fold junctions that result from an intersecting pair of chains.%
\FRAME{ftbpFU}{3.3771in}{2.5443in}{0pt}{\Qcb{Schematic illustration of the
formation of a connected network form disconnected clusters upon increase in
the monomer density $\protect\phi $.}}{\Qlb{percfig}}{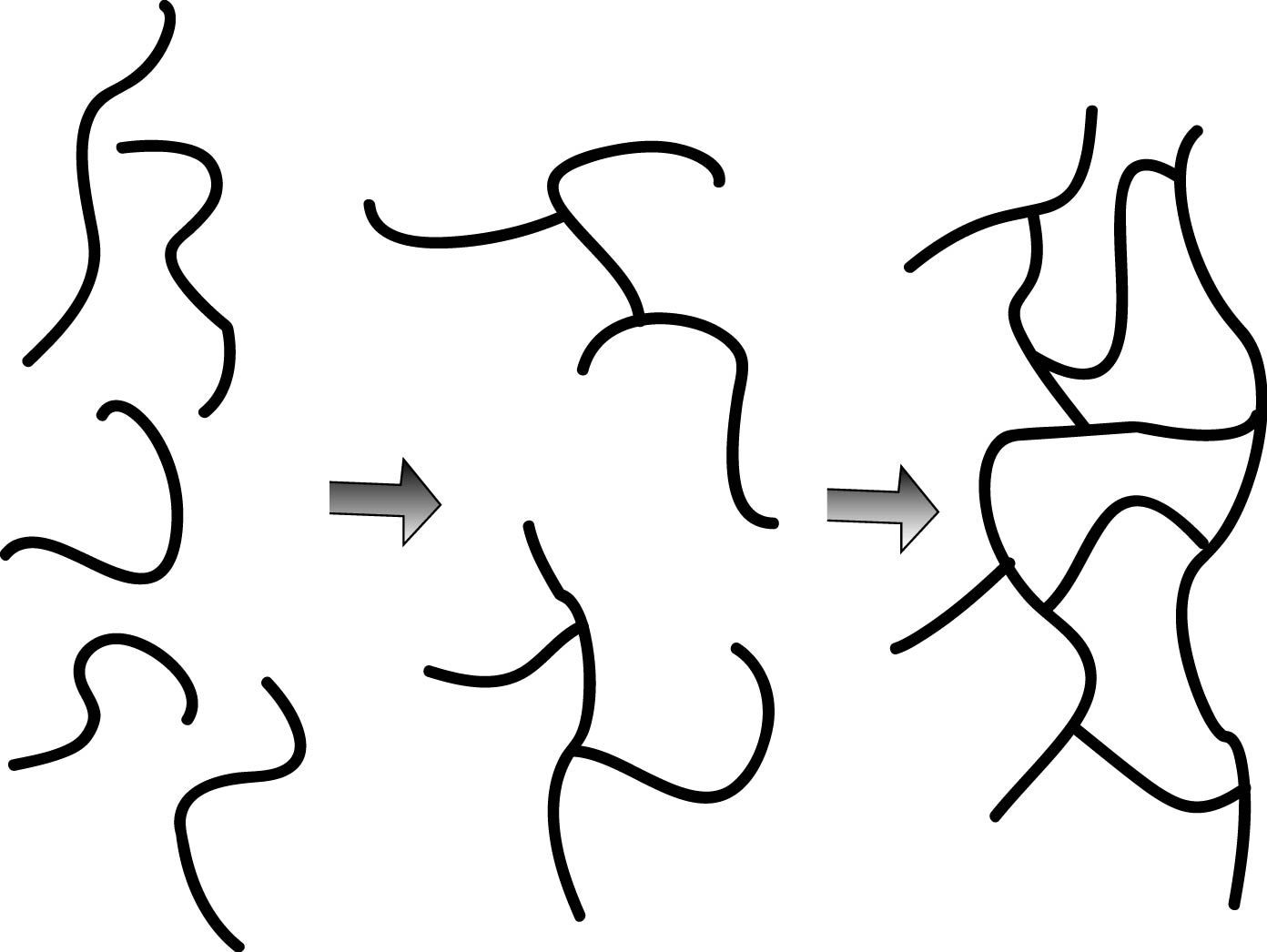}{\special%
{language "Scientific Word";type "GRAPHIC";maintain-aspect-ratio
TRUE;display "USEDEF";valid_file "F";width 3.3771in;height 2.5443in;depth
0pt;original-width 4.6389in;original-height 3.4861in;cropleft "0";croptop
"1";cropright "1";cropbottom "0";filename
'L:/networks/netw.BMP';file-properties "XNPEU";}}

\section{Spatial variations and correlations.\label{corr}}

The advantage of the `$n=0$' model is that it can be used to predict the
spatial correlations of the monomers, ends and junctions, in addition to the
thermodynamic properties discussed above. These correlations are obtained
from calculations of the second moment of the relevant probability
distributions, such as the two-point correlations of the physical
parameters, e.g., the density. The density-density correlation function can
be measured in scattering experiments. Apart from being interesting \textit{%
per se}, the correlations can also provide information about structural,
non-thermodynamic transitions which one might expect in such structurally
complex systems. Here, we generalize the mean field treatment of the
previous section to include the possibility of spatial variations; this
formalism allows us to calculate the correlations in the system. Each spin
is approximated by its ensemble average, but a potential spatial variation
in the value of the average spin (due for example, to a spatially varying
external field)\ is allowed for (cf. Refs.\cite{sambook,crams}). In this
case one must allow for spatial variation of the constants, which are
denoted now as $J_{i}$, $h_{i}$, $K_{i}$ each one labelled by its local
spatial index. Neglecting the fluctuations of each spin about its $local$
average value $\langle S_{i}\rangle $, as in Sec. \ref{mf}, we have 
\begin{equation*}
S_{i}S_{j}=\underset{\text{neglected}}{\underbrace{(S_{i}-\langle
S_{i}\rangle )(S_{j}-\langle S_{j}\rangle )}}-\langle S_{i}\rangle \langle
S_{j}\rangle +\langle S_{j}\rangle S_{i}+\langle S_{i}\rangle S_{j}\simeq
-\langle S_{i}\rangle \langle S_{j}\rangle +\langle S_{j}\rangle
S_{i}+\langle S_{i}\rangle S_{j}
\end{equation*}
where $\langle S_{i}\rangle $ is a $local$ spin average. Similarly we can
approximate the triplet term as: 
\begin{equation*}
S_{i}S_{j}S_{k}\simeq -2\langle S_{i}\rangle \langle S_{j}\rangle \langle
S_{k}\rangle +\langle S_{k}\rangle \langle S_{j}\rangle S_{i}+\langle
S_{i}\rangle \langle S_{k}\rangle S_{j}+\langle S_{i}\rangle \langle
S_{j}\rangle S_{k})
\end{equation*}
Using the same procedure as in Sec. \ref{mf} and writing the average of the
local spin as $\langle S_{i}\rangle \equiv s_{i}$, we find that the
grand-canonical potential per unit volume, $\omega \{s_{i}\}$ (cf. Appendix 
\ref{spinspinapp} for details) is: 
\begin{equation}
\omega \{s_{i}\}V=\frac{1}{2}\sum_{ij}J_{i}s_{i}s_{j}+2\sum_{ijk}\frac{1}{3}%
K_{i}s_{i}s_{j}s_{k}-\sum_{i}\ln (1+\frac{1}{2}\underset{x_{i}}{(\underbrace{%
h_{i}+\sum_{nn_{i}}J_{i}s_{j}+3\sum_{j,k}^{(K)}\frac{1}{3}K_{i}s_{j}s_{k}}}%
)^{2})  \label{omegai}
\end{equation}
where $\underset{jk}{\overset{\left( K\right) }{\sum }}$ is an \emph{%
unconstrained }sum over all possible pairs of sites $j,p$, belonging to the
same triple $i,j,k$. (cf. Appendix \ref{alpha}). 
\begin{eqnarray}
s_{i} &=&-\frac{\partial \omega }{\partial h_{i}}=\frac{x_{i}}{1+\frac{1}{2}%
x_{i}^{2}}  \notag \\
\phi _{i} &=&-\frac{\partial \omega }{\partial \ln J_{i}}=\frac{1}{2}%
\sum_{ij}J_{i}s_{i}s_{j}+\frac{1}{2}h_{i}s_{i}+\frac{3}{2}%
\sum_{ijk}K_{i}s_{i}s_{j}s_{k}=\frac{1}{2}s_{i}x_{i}  \label{phii}
\end{eqnarray}
From this we find: 
\begin{eqnarray}
x_{i} &=&\sqrt{\frac{2\phi _{i}}{1-\phi _{i}}}\text{ and }s_{i}=\sqrt{2\phi
_{i}(1-\phi _{i})}  \label{jixisi} \\
J_{i} &=&\frac{(x_{i}-h_{i}-K_{i}\overset{(K)}{\sum_{\text{\ }jk}}s_{j}s_{k})%
}{\sum_{j}s_{j}}  \notag
\end{eqnarray}
where $\sum_{j}s_{j}$ are sums over nearest neighbors of the site $i$. From
these formulae, the Helmholtz free energy $F\{\phi _{i}\}=\omega V\{\phi
_{i}\}+\sum_{i}\phi _{i}\ln J_{i}$ can be calculated.

For $K_{0}=h_{0}=0$ (corresponding to infinitely long chains) the free
energy is 
\begin{equation}
F\{\phi _{i}\}=\sum_{i}\text{{\LARGE [}}\phi _{i}+\ln (1-\phi _{i})+\frac{1}{%
2}\phi _{i}\ln \frac{2\phi _{i}}{1-\phi _{i}}-\underline{\phi _{i}\ln
\sum_{nn_{i}}\left( 2\phi _{j}(1-\phi _{j})\right) ^{1/2}\text{{\LARGE ]}}}
\label{Fi}
\end{equation}
which we explicitly write here to emphasize the very non-trivial coupling
between the different $\phi _{i}$'s coming from the underlined term. For a
spatially uniform system, $\phi _{i}=\phi ,$ and the $F\{\phi _{i}\}$ has
only the single term $(1-\phi )\ln (1-\phi )$ of Eq.($\ref{FF})$ because the
terms proportional to $1/\bar{N}$ vanish in the limit of $K_{0}=h_{0}=0$
(infinitely long chains) considered here for simplicity.

The focus of the present discussion is different from the treatment of Ref. 
\cite{isaacson}, which was restricted to a study of spin-spin correlations
in the context of the percolation transition. In this section, we focus on
the physically measurable density-density correlations and the thermodynamic
properties. The location of the percolation line can also obtained within
slightly modified formalism, at least at the mean field level, allowing to
relate the thermodynamic properties of the system to its structure.

\subsection{\protect\bigskip Density fluctuations\label{densflucsec}}

Extending the mean-field theory to treat small, local density fluctuations,
the free energy as a function of the local densities, $F\left\{ \phi
_{i}\right\} $ is systematically expanded to quadratic order in the density
fluctuations $\delta \phi _{i},$ related to the local average value of the
density and the mean density by the relation: $\phi _{i}=\phi +\delta \phi
_{i}$. 
\begin{equation}
F\{\phi _{i}\}=F_{MF}(\phi )+\sum_{i}\underset{\frac{\partial F}{\partial
\phi _{i}}}{\underbrace{\ln J_{i}}}\text{ }\delta \phi _{i}+\underset{\delta
^{\left( 2\right) }F}{\underbrace{\sum_{ki}S_{ki}^{-1}\delta \phi _{k}\delta
\phi _{i}}}+O(\delta \phi ^{3})  \label{fij}
\end{equation}
where $F_{\text{MF}}(\phi )$ is just the mean field free energy of Sec. \ref
{mf}; $F_{\text{MF}}(\phi )=Vf_{\text{MF}}(\phi )$, $S_{ik}^{-1}=\frac{%
\partial ^{2}F\{\phi _{i}\}}{\partial \phi _{i}\partial \phi _{k}}$ is a
matrix of the second derivatives of $F\{\phi _{i}\}$, and we have used the
fact that $\frac{\partial F}{\partial \phi _{i}}=\mu _{i}=\ln J_{i}$.
Expansion in Eq.$\left( \ref{fij}\right) $ is a completely general one, and
the particular expressions for $S_{ik}^{-1}$ can be obtained from Eqs. (\ref
{jixisi},\ref{phii}). By the definition of the constant $J_{i},$ as an
exponent of a local chemical potential, the density-density fluctuation
function is given by $\langle \phi _{i}\phi _{k}\rangle -\phi ^{2}=\frac{%
\partial ^{2}\ln Z\{\phi _{i}\}}{\partial \ln J_{i}\partial \ln J_{k}}$
which turns out to be\emph{\ }the inverse of the matrix\ $S_{ik}^{-1}$: 
\begin{equation*}
\langle \phi _{i}\phi _{k}\rangle -\phi ^{2}=\frac{1}{V}\frac{\partial \phi
_{i}}{\partial \ln J_{k}}=\frac{1}{V}S_{ik}
\end{equation*}
where we have used the fact that $\frac{\partial \ln Z\{\phi _{i}\}}{%
\partial \ln J_{i}}=\phi _{i}$. A cumbersome but straightforward calculation
using the Eq.(\ref{jixisi})(cf. Appendix \ref{dens-dens} ) gives

\begin{eqnarray}
\frac{\partial ^{2}F\{\phi _{i}\}}{\partial \phi _{i}\partial \phi _{k}} &=&%
\frac{\partial \ln J_{i}}{\partial \phi _{k}}=\frac{1}{2\phi (1-\phi )}%
(\delta _{ik}-(1-2\phi )\frac{1}{q}\sum_{j=nn_{i}}\delta _{jk})+  \notag \\
&&+\frac{h_{0}}{2q^{1/2}(2\phi )^{3/2}}\frac{1}{(1-\phi )}(\delta
_{ik}+(1-2\phi )\frac{1}{q}\sum_{j=nn_{i}}\delta _{jk})  \label{sij1} \\
&&-\frac{3K_{0}(\alpha ^{\prime }/3)}{q^{3/2}(2\phi )^{1/2}}\frac{1}{(1-\phi
)}(\frac{1}{2}\delta _{ik}-(1-2\phi )\frac{1}{6q}\sum_{j=nn_{i}}\delta _{jk}+%
\frac{2}{\alpha ^{\prime }}(1-2\phi )\sum_{j=nnn_{i}}\delta _{jk})  \notag
\end{eqnarray}
Writing the Fourier transform 
\begin{equation*}
\delta \phi _{i}=\sum_{\vec{p}}\delta \phi (\vec{p})e^{i\vec{p}\cdot \vec{r}%
_{i}}
\end{equation*}
we get 
\begin{equation*}
\delta ^{\left( 2\right) }F=\sum_{\vec{p}}S^{-1}(\vec{p})\delta \phi (\vec{p}%
)\delta \phi (-\vec{p})\text{ and }\langle \delta \phi (\vec{p})\delta \phi
(-\vec{p})\rangle =\frac{1}{V}S(\vec{p})\text{\ }
\end{equation*}
with $S(\vec{p})=\underset{ik}{\sum }$ $S_{ik}e^{i\vec{p}(\vec{r}_{i}-\vec{r}%
_{k})}$. From Eq.$\left( \ref{sij1}\right) $\ we find (for the simple cubic
lattice) 
\begin{eqnarray}
S^{-1}(\vec{p}) &=&\frac{1}{2\phi (1-\phi )}[A+B(\cos (p_{x})+\cos
(p_{y})+\cos (p_{z}))+  \label{Sp} \\
&&+C(\cos (p_{x})\cos (p_{y})+\cos (p_{z})\cos (p_{y})+\cos (p_{x})\cos
(p_{z}))]  \notag
\end{eqnarray}
where 
\begin{eqnarray*}
A &=&1+\frac{h_{0}}{2(2\phi q)^{1/2}}-\frac{3K_{0}(\alpha ^{\prime }/3)}{%
2q^{3/2}}(2\phi )^{1/2} \\
B &=&(2\phi -1)[\frac{2}{q}-\frac{h_{0}}{(2\phi )^{1/2}q^{3/2}}-\frac{%
3K_{0}(\alpha \prime /3)}{3q^{5/2}}(2\phi )^{1/2}] \\
C &=&(2\phi -1)\frac{4K_{0}(\alpha \prime /3)}{q^{5/2}}(2\phi )^{1/2}
\end{eqnarray*}
and we have used the fact that the Fourier transform ($FT)$ of the delta
function is equal to unity and 
\begin{equation*}
FT[\sum_{j}\delta _{ij}]=\sum_{\alpha }e^{ip_{\alpha }\hat{e}_{\alpha }}
\end{equation*}
where $\hat{e}_{\alpha }$ is the unit vector that points in the direction of
the site $j$. On a simple cubic lattice, $p_{\alpha }$ is the component of
the vector $\vec{p}$ in the direction $\alpha .$ As one can easily see from
Eqs.$\left( \ref{FF},\ref{Sp}\right) $, the inverse of the structure factor
at the zero wavevector $S\left( 0\right) =\int S(r)dr$ is equal to the
second derivative of the \emph{mean field }free energy: 
\begin{equation}
S^{-1}(0)=\frac{\partial ^{2}F_{\text{MF}}(\phi )}{\partial \phi ^{2}}=\frac{%
1}{1-\phi }+\frac{h_{0}}{(2\phi )^{3/2}q^{1/2}}-\frac{3K_{0}\left( \alpha
\prime /3\right) }{(2\phi )^{1/2}q^{3/2}}  \label{s0}
\end{equation}
This is an expression of the fluctuation-dissipation theorem in the
grand-canonical ensemble. The variance of the mean number of monomers in the
grand canonical ensemble is 
\begin{equation}
\langle N^{2}\rangle -\langle N\rangle ^{2}=\int dr\langle (\phi (r)-\phi
(0))^{2}\rangle =TV\phi ^{3}\kappa _{T}=S(0)  \label{kappat}
\end{equation}
where $\kappa _{T}\,$\ is the isothermal compressibility\cite{huang}. Noting
that 
\begin{equation*}
\frac{\partial ^{2}F_{\text{MF}}(\phi )}{\partial \phi ^{2}}=\frac{\partial
\mu }{\partial \phi }=\frac{1}{\phi ^{3}}\frac{\partial \Pi }{\partial v}=%
\frac{1}{\phi ^{3}}\kappa _{T}^{-1}
\end{equation*}
($\Pi $ is the osmotic pressure, $v=1/\phi $ and $\mu =\partial F_{\text{MF}%
}(\phi )/\partial \phi $\ is the chemical potential), Eq.$\left( \ref{s0}%
\right) $ becomes identical to Eq.$\left( \ref{kappat}\right) .$ Thus, the
calculated density-density correlation function satisfies the expected
thermodynamic sum rules.

The structure factor $S\left( \vec{p}\right) =\langle \delta \phi (\vec{p}%
)\delta \phi (-\vec{p})\rangle $ is experimentally measurable by scattering
experiments (neutron, light, X-ray etc.) where the intensity of scattered
radiation at the wavevector $\vec{p}$ is proportional to $S\left( \vec{p}%
\right) .$ If $S(0)$ is determined experimentally as a function of $\phi ,$
the free energy $F_{\text{MF}}(\phi )$ can be found by the integration of Eq.%
$\left( \ref{s0}\right) .$ Obviously, Eq.$\left( \ref{Sp}\right) $ is not
valid for the values of parameters at which the phase separation is
observed, that is, in the region where $\frac{\partial ^{2}F(\phi )}{%
\partial \phi ^{2}}<0$. Rather, each macroscopic phase in a two-phase
equilibrium has its own value of $S(\vec{p}),$ with the value of the monomer
density appropriate for each phase.

The expression (\ref{Sp}) for $S(\vec{p})$ is quite complicated, and its
behavior is rather different for low and high densities. We next consider
the nature of the predicted structure factor of Eq.$\left( \ref{Sp}\right) $
in the limits of high and low monomer density, $\phi .$

\subsubsection{Low density limit\label{low}}

For small densities $\phi $ the coefficients $B$ and $C$ are negative, and
one can expand Eq.$\left( \ref{Sp}\right) $ in a power series in the
wavevector $\vec{p}$. One finds that the coefficient of $p^{2}$ is positive
so that this expansion is reasonable for wavevectors that are small compared
with the inverse of the monomer size. 
\begin{eqnarray}
S^{-1}(\vec{p}) &\simeq &S^{-1}(0)+ap^{2}\text{ so that \ }S(\vec{p})\simeq 
\frac{S(0)}{1+aS(0)p^{2}}\text{ }  \label{smallfi} \\
S(0) &=&(\frac{\partial ^{2}F_{\text{MF}}(\phi )}{\partial \phi ^{2}})^{-1}%
\text{ and \ }a=\frac{(1-2\phi )}{2\phi (1-\phi )}(1+\frac{3K_{0}\left(
\alpha ^{\prime }/3\right) }{q^{3/2}}\frac{15}{4}(2\phi )^{1/2}-\frac{h_{0}}{%
(2\phi q)^{1/2}})  \notag
\end{eqnarray}
Note that $a$ is a non-negative, because $\frac{h_{0}}{(2\phi q)^{1/2}}=%
\frac{\phi _{e}}{2\phi },$ the half of the ratio of the number of ends to
the number of junctions. In the present approximation of sparse junctions
and ends,$\frac{\phi _{e}}{\phi }\ll 1$ and $a$ is non-negative. In the case
of linear polymers without junctions, $\frac{\phi _{e}}{2\phi }$ is equal to
the inverse average length of the chains, $\frac{\phi _{e}}{2\phi }=\frac{1}{%
\bar{N}}$, which leads to the known correction to the scattering structure
factor \cite{doibook}.

In real space, the density-density correlation function is given by Fourier
transform of the $S\left( \vec{p}\right) $. It follows from Eq.$\left( \ref
{smallfi}\right) $ that in real space $\langle \delta \phi (\vec{r})\delta
\phi (0)\rangle $ has the usual Ornstein-Zernicke form

\begin{equation*}
\langle \delta \phi (\vec{r})\delta \phi (0)\rangle =\sum_{\vec{p}}S(\vec{p}%
)e^{i\vec{p}\cdot \vec{r}}\simeq \frac{1}{r}e^{-r/\xi }
\end{equation*}
with the correlation length $\xi $ given by 
\begin{equation}
\xi =a\left( \frac{\partial ^{2}F_{\text{MF}}(\phi )}{\partial \phi ^{2}}%
\right) ^{-1}  \label{xi}
\end{equation}
The correlation length of density fluctuations, $\xi ,$ diverges at the
spinodal line of the first-order junction-ends transition studied in the Sec.%
\ref{j-e}, at which $\frac{\partial ^{2}F_{\text{MF}}(\phi )}{\partial \phi
^{2}}=0$; this is expected for any first-order transition. Similarly, from
Eq.$\left( \ref{smallfi}\right) $ we see that the density fluctuations at
zero wavevector$,$ $S$($\vec{p}=0)$, are also divergent at the spinodal,
and, in particular, at the critical point. The divergence of the scattering
intensity at the critical point should be observable in scattering
experiments in the same way as the usual critical opalescence. We note in
passing that in the absence of junctions, the structure factor, as given by
Eq.$\left( \ref{smallfi}\right) ,$ reduces to the classical RPA result for
polymer solutions \cite{doibook}.

It is interesting to note that the \emph{overall} structure of a solution of
self-assembled branched aggregates is not strictly self-similar, as we can
see from Eq.$\left( \ref{smallfi}\right) .$ There is at least one
characteristic length in the problem, namely, the correlation length $\xi .$
At distances larger than $\xi $ the correlations decay exponentially while
for self-similar structures one expects algebraic decay of correlations.
However, each branched cluster is expected to be self-similar, in analogy to
percolation clusters\cite{perc1}. Note that both the correlation length $\xi 
$ and the scattering intensity at zero wavevector $S\left( 0\right) $ are
both proportional to the second derivative of the free energy,$\frac{%
\partial ^{2}F_{\text{MF}}\left( \phi \right) }{\partial \phi ^{2}}.$ This
means that both of them are \emph{non-monotonic} functions of the density,
but have a maximum around the line $\phi _{e}\approx \phi _{j}$, as follows
from Eqs.$\left( \ref{secder}\right) \left( \ref{xi}\right) $, by taking the
derivatives $\frac{\partial \xi }{\partial \phi }$, $\frac{\partial S\left(
0\right) }{\partial \phi }$. The maxima of $S\left( 0\right) $ and of $\xi $
as a function of $\phi $ are determined by the condition $\frac{\partial
S\left( 0\right) }{\partial \phi }=\frac{1}{S\left( 0\right) ^{2}}\frac{%
\partial ^{3}F_{\text{MF}}\left( \phi \right) }{\partial \phi ^{3}}=0$,
which is $precisely$ the same condition which determines the location of the
critical point ($\frac{\partial ^{3}F_{\text{MF}}\left( \phi \right) }{%
\partial \phi ^{3}}|_{\phi _{c}}=0)$. In particular, this leads to the
conclusion that the maximum scattering in the region above the phase
separation should be observed along a line which starts at the critical
point and is determined by the condition $\frac{\partial ^{3}F_{\text{MF}%
}\left( \phi \right) }{\partial \phi ^{3}}=0$ ( equivalent for the small
densities to the condition $\phi _{e}\approx \phi _{j})$ , as shown in Fig. 
\ref{corrfig}. This is in agreement with observations from scattering
experiments from solutions of branched wormlike micelles \cite{khatory}(cf.
also Discussion).

It is instructive to compare $\xi $ with another length scale that is
present in the problem, namely, $\bar{l}$, the mean distance between the
defects (i.e. ends or junctions). As shown in Sec.\ref{mf}, 
\begin{equation}
\bar{l}=\frac{2\phi }{3\phi _{j}+\phi _{e}}\simeq \frac{\phi }{%
3K_{0}^{\prime }\phi ^{3/2}+h_{0}^{\prime }\phi ^{1/2}}  \label{lbarxi}
\end{equation}
and $\bar{l}(\phi )$ has a maximum around $\phi \sim h_{0}^{\prime
}/K_{0}^{\prime }$. The behavior of $\bar{l}$ is quite different from that
of $\xi ,$ indicating that \emph{these two lengths are physically unrelated}$%
.$ The behavior of $\bar{l}$ is non-singular, while the correlation length $%
\xi $ diverges at the spinodal line of the junction-ends transition. This
emphasizes the fact that the junction-induced attraction has an effect on
all the monomers in the system, as reflected in the behavior of $\xi ;$ the
effect of junction induced attraction is not limited to the behavior of the
junctions alone; this is reflected in the non-singular nature of $\bar{l}.$
The physical meaning of $\xi $ is similar to the `blob' size of the
semidilute polymer solutions. Namely at distances smaller than $\xi $, the
behavior of an aggregate is that of a single self-avoiding branched polymer.

\subsubsection{High densities\label{peak}}

\bigskip When $\phi >1/2,$ the coefficient ($2\phi -1)$ in Eq.$\left( \ref
{Sp}\right) $ becomes positive, which results in a negative coefficient in
front of the term proportional to $p^{2}$ (in the expansion of the structure
factor for small wavevectors). Thus, the small wavevector \ expansion
becomes meaningless, since it would give a negative structure factor for
high enough values of $p$. Therefore, in this region one cannot use the
expansion and one must retain the full form of the $S(\vec{p}).$ The exact
location of the line where the behavior changes, $\phi =1/2,$ is an artifact
of the lattice model and should not be taken literally. Inspection of Eq.$%
\left( \ref{Sp}\right) $ reveals that for small enough junctions densities,
( small Boltzmann factor $K_{0}=e^{-\epsilon _{j}/T}),$ \ for which 
\begin{equation*}
|B/(2C)|<1\Rightarrow K_{0}^{\prime }<\frac{4}{17}\frac{(1-\frac{1}{2}\frac{%
h_{0}}{(2\phi q)^{1/2}})}{(2\phi )^{1/2}}\equiv K^{\star }
\end{equation*}
($K_{0}^{\prime }=K_{0}\alpha /q^{3/2})$, the structure factor $S\left( \vec{%
p}\right) $ is a monotonically increasing function of the wavevector, $p$
(for concreteness we consider the [1,1,1] direction) for all $\phi >1/2.$

This behavior of the structure factor at small junctions densities (weakly
branched chains) agrees with the previous studies of the density
fluctuations in concentrated polymer solutions and melts \cite{curroexp}.
Indeed, it is known that the structure factor of concentrated polymer
solutions and melts behaves qualitatively like that of simple liquids\cite
{curroexp}. In particular, for small wavevectors $p$, the $S\left( p\right) $
is an increasing function of \ $p$. This is related to the low
compressibility of dense liquids. As discussed in the previous section, as
the monomer density $\phi $ tends to one, the compressibility tends to zero.
The magnitude of the density fluctuations, related to the compressibility by
the fluctuation-dissipation theorem tends to zero as well \cite{huang}. \
When the monomer density $\phi $ is high, but \ lower than one, the long
wavelength (small $p)$ density fluctuations are small, because large
collective rearrangements of the monomers are structurally inhibited.
However, short wavelength, local, rearrangements of the molecules are still
possible. Combined together, these two effects produce a structure factor
that increases with increasing wavevector, $p$.

In real systems, for the values of $p$ larger than the inverse molecular
size, one observes oscillations of the structure factor as a function of $p$%
, with the peaks corresponding to the first-shell, second-shell, etc.,
correlations of the positions of the adjacent monomers. Although the lattice
model in the mean field approximation cannot be used to predict precisely
the extremely low short scale features, such as the first shell peak
observed in scattering experiments on polymer solutions, it adequately
treats the qualitative behavior of the structure factor for wavevectors
smaller than the inverse of the lattice constant.

For higher values of $K_{0}$, $K^{\prime }>K^{\star },($i.e. for higher
junction densities) a $peak$ emerges in $S\left( \vec{p}\right) ($ see Fig. 
\ref{corrfig}) at a value of wavevector given by: 
\begin{equation}
p_{0}=\arccos \text{{\Large [}}-\frac{B}{2C}\text{{\Large ]}}  \label{p00}
\end{equation}
indicating the presence of medium range correlations. This peak is quite
different form the first-shell peak of simple fluids, which reflects
excluded volume interactions on the molecular scale. On the contrary, the
peak of Eq.$\left( \ref{p00}\right) $ reflects medium range structural
correlations. By medium range we mean that the correlation range is larger
than the lattice constant. The location of the peak depends on the values of
the parameters, as discussed below, and it is not a lattice artifact.
Because the peak appears only deep in the network phase, where the number of
ends is negligible compared to the number of the junctions, one can neglect
the $h_{0}$-dependent term in Eq.(\ref{Sp}) as a first approximation and
write: 
\begin{equation}
p_{0}\cong \arccos \text{{\Large [}}\frac{1}{4}(1-\frac{1}{4K_{0}^{\prime
}(2\phi )^{1/2}})\text{{\Large ]}}  \label{p0}
\end{equation}
indicating that the location of the peak moves towards lower values of $p$
as $K_{0}^{\prime }\left( 2\phi \right) ^{1/2}$ increases. Fig. \ref{peakfig}
shows the structure factor for two values of $K_{0}^{\prime }$: below and
above $K^{\star }$. Although Eq.$\left( \ref{p0}\right) $ was obtained for
the [1,1,1] direction ($p_{x}=p_{y}=p_{x})$, a similar result would be
obtained for the isotropically averaged structure factor $\bar{S}\left(
p\right) =\int d^{3}pS\left( \vec{p}\right) $. The peak in $\bar{S}\left(
p\right) $ should appear whenever there is a peak in any `crystallographic'
direction. In isotropic systems, the averaged structure factor $\bar{S}(p)$
is, of course, the experimentally measurable quantity, free of lattice and
model artifacts. The predicted peak in the structure factor due to the
presence of the junctions, is not a lattice artefact, because the location
of this peak location is not related to the lattice constant, but determined
solely by the numbers of junctions present in the system.

The precise size of the junction is arbitrary to a certain extent and can be
adjusted to reflect the real physical features of any given system. We have
defined a junction as a connection between three adjacent sites. However,
one can also consider a larger junction, connecting next nearest neighbors,
as in Fig. \ref{hexfig}(b). Intuitively, one expects a more pronounced peak
for larger junctions, because it amounts to an increase of the junction
volume. Exact numerical calculations show that it is indeed the case.

In the context of microemulsions, this peak can be related to the
experimentally observed peak in the scattering from bicontinuous
microemulsions \cite{gommpershik,strey} (cf. Discussion). It is important to
realize that the absence of a peak \textit{does not}, in principle, imply
the absence of a network, (cf. Fig. \ref{corrfig}).

\FRAME{ftbpFU}{3.3935in}{2.3851in}{0pt}{\Qcb{Structure factor as a function
of wavevector $p$ in the high density region,$\protect\phi >\frac{1}{2}.$
Dashed line shows the structure factor at low tempertaures, where the number
of thermally-generated junctions is small. For higher temperatures, a peak
developes in the structure factor, shown by the thick line.}}{\Qlb{peakfig}}{%
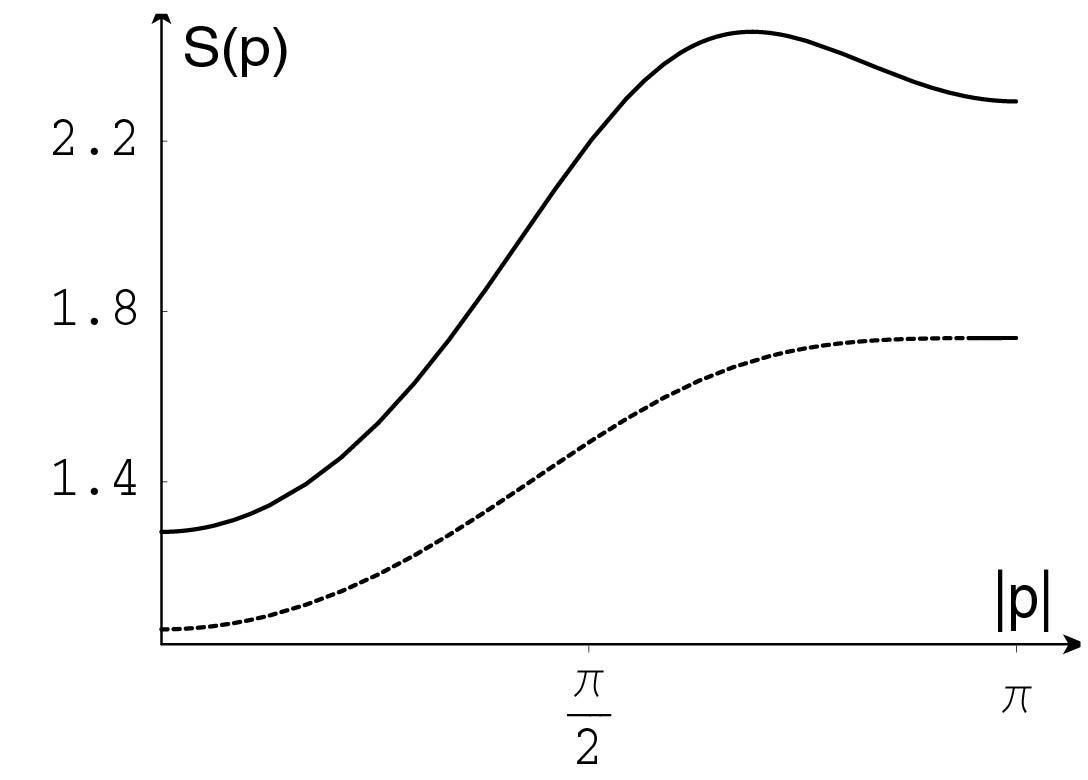}{\special{language "Scientific Word";type "GRAPHIC";display
"USEDEF";valid_file "F";width 3.3935in;height 2.3851in;depth
0pt;original-width 22.6667in;original-height 16.2395in;cropleft "0";croptop
"1";cropright "1";cropbottom "0";filename
'L:/networks/peak.jpg';file-properties "NPEU";}}

\section{Spin-spin correlations and topological structure.\label{spinspin}}

So far we have been concerned with thermodynamic properties of a system and
related phenomena such as equilibrium density fluctuations. They can be
adequately described by a model presented in Sec. \ref{juncn0} which maps
the solution of equilibrium branched clusters onto Heisenberg model with
anisotropic $S_{i,1}S_{j,1}S_{k,1}$ three-spin term. However, no structural
information, e.g., concerning the formation of a continuous network, can be
extracted from the model as it is formulated in Sec. \ref{juncn0}.
Unfortunately, in order to be able to extract structural information about
the network topology from the model, \textit{while retaining the exact
correspondence }between the spin model and the physical system of branched
clusters\textit{, }one has to resort to tensor order parameter, which
results in a formally complicated theory \cite{isaacson}. Instead, we shall
use a simpler, but less rigorous theory, applicable \textit{only }in the
mean field approximation, employed in this paper.

Although the spin $\vec{S}$ has `zero' components, one can still think about
a `parallel' component $S_{1}$ pointing in the direction of the field $\vec{h%
}$ and a `transverse' component $S_{\bot }.$ The rigorous mathematical
procedure that justifies this is shown in Appendix. \ref{spinspinapp} and
amounts to calculating all the quantities of interest for a finite number of
components $n$, and taking the limit $n\rightarrow 0$ subsequently. One can
then devise a slightly more generalized version of the Hamiltonian \ref{h3},
where the three-spin term takes the following form: 
\begin{equation}
K\sum_{ijk}\left[ S_{i,1}S_{j,1}S_{k,1}+\frac{\beta }{3}\left(
S_{i,1}(S_{j,\bot }\cdot S_{k,\bot })+S_{j,1}(S_{i,\bot }\cdot S_{k,\bot
})+S_{k,1}(S_{i,\bot }\cdot S_{j,\bot })\right) \right]  \label{beta}
\end{equation}
where the sum is over distinct triples $ijk$ and $\beta $ is a numerical
parameter. One can also include coupling between the 'transverse' components
of the spin $\vec{S}$, consistent with the symmetries of the system. This
modification of the Hamiltonian has no effect on the thermodynamic
properties and equilibrium fluctuations \textit{in the mean field
approximation}, discussed so far. Calculation of either the partition
function or the two-point correlation function has contributions from the
terms proportional to $\beta $ that have at least one power of the average
of the transverse component $\langle S_{\bot }\rangle .$ Since $\langle
S_{\bot }\rangle =0,$ terms proportional to $\beta $ \textit{do not}
contribute to thermodynamic quantities at the level of the mean field
approximation. For $\beta =0$ the Hamiltonian $\left( \ref{beta}\right) $
reduces to the previously used one. For $\beta \neq 0$ the correspondence
between the spin model and the equilibrium branched clusters $\ is$ $not$ $%
exact$. The clusters containing internal loops enter into expansion with the
wrong weight due to summation over $n-1$ 'transverse' components. However,
with the choice $\beta =3$ all $loopless$ tree-like clusters are counted
correctly (cf. Appendix \ref{spinspinapp}). Mean field approximation
neglects the presence of intra-cluster loops, which formally correspond to $%
correlations$ between different spins. This can be seen from the fact that
the mean field free energy does not depend on the value of $\beta $ in Eq.$%
\left( \ref{beta}\right) .$ That is, it is independent of the precise way in
which weights are assigned to the internal loops, which means that the intra
cluster loops are neglected in the mean field approximation (cf. also
Appendix \ref{spinspinapp}). Thus, at the mean field level, the Eq.$\left( 
\ref{beta}\right) $ with $\beta =3$ can be used to calculate the transverse
spin correlations which are related to the cluster size and percolation
threshold. \ Therefore,\ we use the $\beta =3$ model which neglects
inter-cluster long-range correlations, responsible for the loops,
consistently with the mean field approximation.

Although the spin $\vec{S}$ itself has no physical meaning, it enters the
calculation of the various physical quantities, as we have seen in Sec. \ref
{mf}. Analogously, the spin-spin correlation functions $\langle
S_{i,a}S_{j,a}\rangle $, although unphysical themselves, can be used to
derive physically relevant quantities such as the density-density
correlation function. Another interesting, physical observable (e.g., by
scattering experiments conducted in solutions of end-labelled chains) is the
correlation between the end-points of the aggregates, which is related to
the spin-spin correlation function by (cf. Appendix. \ref{spinspinapp}) 
\begin{equation}
\langle \phi _{e}(r_{i})\phi _{e}(r_{j})\rangle -\phi _{e}^{2}=\frac{%
\partial h_{i}\langle S_{i}\rangle }{\partial h_{j}}=\delta _{ij}+h[\langle
S_{i}S_{j}\rangle -\langle S_{i}\rangle \langle S_{j}\rangle ]  \label{endsr}
\end{equation}
This can be understood geometrically by noting that the graphs which enter
into $\langle S_{i}S_{j}\rangle $ are those with ends present both at the
site $i$ and site $j.$(cf. Fig. \ref{spinfig} and Appendix \ref{spinspinapp}%
).

Similarly, although $S_{\bot }$ itself does not have a direct physical
interpretation, its correlator $\langle S_{i,\bot }S_{k,\bot }\rangle $
does. If one considers the expansion of the correlation function $\langle
S_{i,\bot }S_{k,\bot }\rangle $ (cf. Appendix \ref{spinspinapp}) in powers
of $J$, $h$ ( in the absence of junctions, i.e., $K=0)$ one can see that the
only non-zero contributions arise from configurations that contain a single
chain which starts at the site $i$ and ends at the site $k$. Consequently, $%
\langle S_{i,\bot }S_{k,\bot }\rangle $ measures the correlations between
the ends of a single chain. As a matter of fact, even when $K\neq 0$, $%
\langle S_{i,\bot }S_{k,\bot }\rangle $ is non-zero only when $i$ and $k$
are the ends belonging to the $same$ cluster (cf. Appendix \ref{spinspinapp}
and Fig. \ref{spinfig}). Thus, in general, the \textit{`transverse'}
correlation function measures the correlations between chain ends belonging
to the \textit{same} aggregate. Similarly the \textit{`longitudinal'}
correlation function $\langle S_{i,1}S_{k,1}\rangle $ measures the end-end
correlations between \textit{any two ends}, even if they belong to different
chains or aggregates. The density-density correlations can be calculated
from the `longitudinal' correlator $\langle S_{i,1}S_{k,1}\rangle $ using
Eq. (\ref{phii}) that relates $\phi _{i}$ to $S_{i}.$ This calculation
yields result identical to those obtained in Sec. \ref{dens-dens}.

In this section, we present the results of a calculation of the spin-spin
correlators $\langle S_{i,a}S_{j,a}\rangle $ using the same local mean field
approximation which was used in Sec. \ref{mf} to find the density-density
correlation function. Each spin is approximated by its local ensemble
average: $s_{i,\alpha }=\langle S_{i,\alpha }\rangle _{local}$ (with $%
a=\{1,\bot \}$), which can, however, be different for spins at\emph{\ }%
different sites labelled by $i.$ Eventually, we will be interested in the
deviations $\delta s_{i}$ of the \textit{local} average spins from the 
\textit{spatially} averaged value $S.$ The full calculation is presented in
Appendix. \ref{spinspinapp}. The spin-spin correlator is obtained from the
partition function $Z$ using the relations:

\begin{equation*}
\langle \delta s_{i,a}\delta s_{j,a}\rangle =\frac{\partial ^{2}}{\partial
h_{i}\partial h_{j}}\ln Z=\frac{\partial s_{i,a}}{\partial h_{j,a}}
\end{equation*}
Using the results of Eq.$\left( \ref{phii}\right) ,$ a tedious but
straightforward calculation yields the Fourier transform of the spin-spin
correlator, $C_{\vec{p}}^{a}=\sum_{\vec{p}}e^{i\vec{p}(\vec{r}_{i}-\vec{r}%
_{k})}\langle \delta s_{i,a}\delta s_{j,a}\rangle :$ 
\begin{equation}
C_{\vec{p}}^{a}=\frac{A^{a}}{1-A^{a}(Jg_{1}(\vec{p})+6Ksg_{2}(\vec{p}))}
\label{cpa}
\end{equation}
where (cf. Appendix. \ref{spinspinapp}) 
\begin{eqnarray*}
A^{(1)} &=&(1-2\phi )(1-\phi ) \\
A^{\left( \bot \right) } &=&(1-\phi )
\end{eqnarray*}
and where 
\begin{eqnarray*}
\text{ }g_{2}(\vec{p}) &=&\frac{2}{3}\frac{\alpha }{q}\sum_{\text{unit cell}%
}e^{i\vec{p}\hat{e}_{\alpha }}+\sum_{\text{second cell}}e^{i\vec{p}\hat{e}%
_{\alpha }} \\
g_{1}(\vec{p}) &=&\sum_{\text{unit cell}}e^{i\vec{p}\hat{e}_{\alpha }}
\end{eqnarray*}
Consequently, the spin-spin correlation function in Fourier space is: 
\begin{equation*}
\langle \delta s_{a}(\vec{p})\delta s_{a}(\vec{p}^{\prime })\rangle =C_{\vec{%
p}}^{a}\delta _{\vec{p},-\vec{p}^{\prime }}
\end{equation*}
In the following we show these results can be used to determine various
physical characteristics of the system.

\FRAME{ftbpFU}{3.3771in}{2.29in}{0pt}{\Qcb{At low densities, $\protect\phi <%
\frac{1}{2}$, the structure \ is a decreasing function of thewavevector. The
scattering intensity at zero wavevector and the correlation length diverge
at the phase stability coundary of the junctons-ends transition, shown by
the thick closed line. In the region above the phase separation, the
scattering intensity has a maximum along the dashed line, starting at the
critical point. For high densities and low numbers of junctions, the
structure factor is an increasing function of the wavevector, while for
higher numbers of junctions present, a peak develops in the structure factor}%
}{\Qlb{corrfig}}{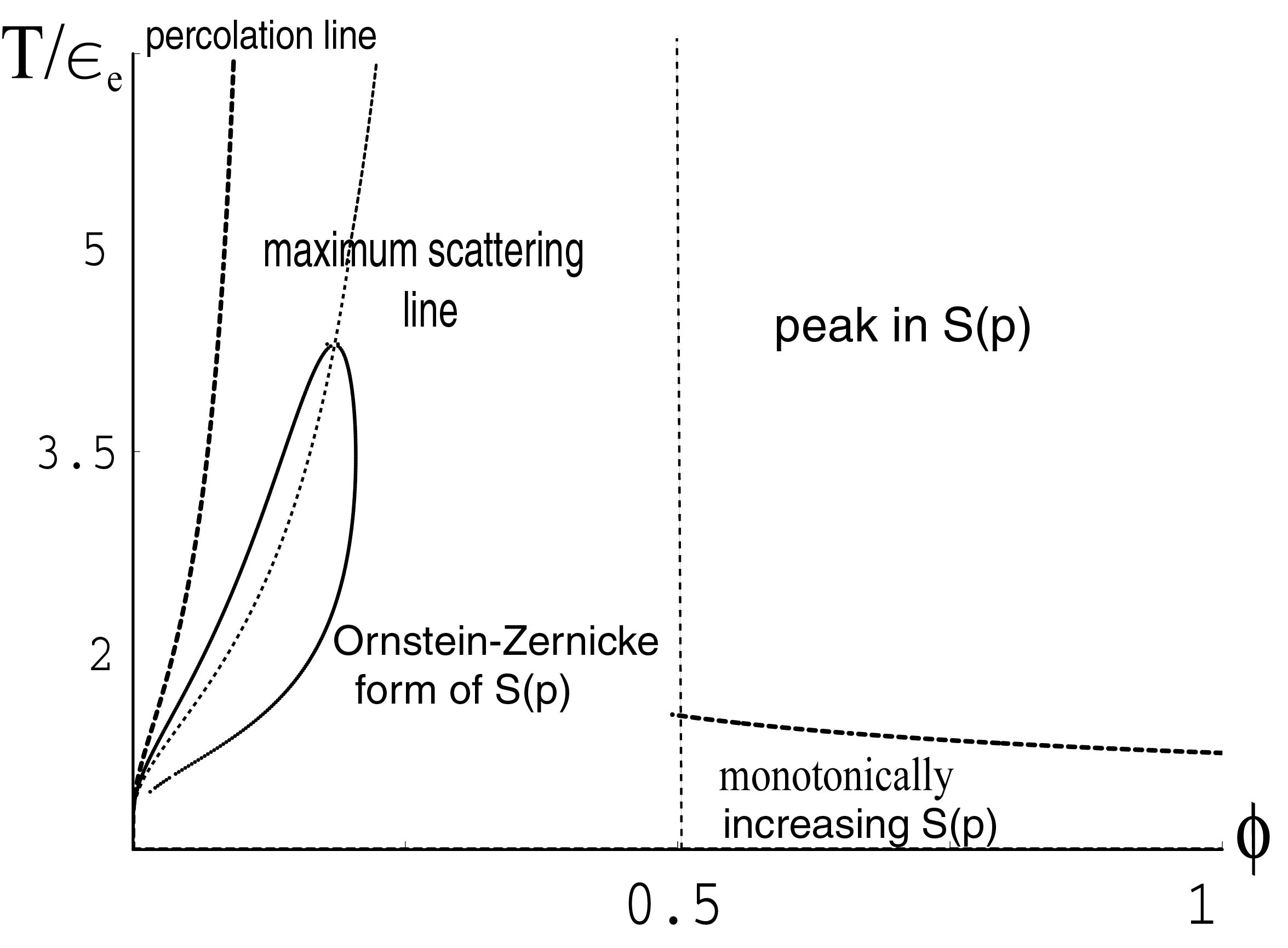}{\special{language "Scientific Word";type
"GRAPHIC";display "USEDEF";valid_file "F";width 3.3771in;height 2.29in;depth
0pt;original-width 24.4898in;original-height 18.021in;cropleft "0";croptop
"1";cropright "1";cropbottom "0";filename
'L:/networks/corr.jpg';file-properties "NPEU";}}

\subsection{Longitudinal correlations and the defect correlation function%
\label{longitudinal}}

\bigskip The correlation function of the aggregate ends can be calculated
form Eq.$\left( \ref{cpa}\right) $, giving in Fourier space 
\begin{eqnarray}
\langle \delta \phi _{e}(\vec{p})\delta \phi _{e}(-\vec{p})\rangle &=&1+hC_{%
\vec{p}}^{1}=  \notag \\
\langle \delta \phi _{e}(\vec{p})\delta \phi _{e}(-\vec{p})\rangle -1
&=&h(1-2\phi )S(\vec{p})  \label{endsp}
\end{eqnarray}
where $S\left( \vec{p}\right) $ is the same as in Eq.$\left( \ref{Sp}\right) 
$ for density-density correlations. Apart from the prefactor and the
constant ( equal to $1)$ that ensures that the correlation of an end with
itself is not counted, the dependence of $\langle \delta \phi _{e}(\vec{p}%
)\delta \phi _{e}(-\vec{p})\rangle $ on the wavevector $p$ is identical to
that of the full density-density correlation function of Sec. \ref{dens-dens}%
\emph{.} This is not accidental, but rather is a direct consequence of the
fact that the density-density correlation function is related to spin-spin
correlation function via Eq.(\ref{phii}). In particular, the correlation
length at small densities $\xi $ is the same as the correlation length\ of
the total density fluctuations. The same is true of\emph{\ }the
junction-junction correlation function 
\begin{equation*}
\langle \phi _{j}(r_{i})\phi _{j}(r_{j})\rangle -\phi _{j}^{2}=\frac{%
\partial K_{i}\sum_{i,k,m}s_{i}s_{k}s_{m}}{\partial K_{j}}=\delta
_{ij}K\alpha s^{3}+6Ks^{2}\sum_{p,j}^{(K)}\frac{\partial s_{p}}{\partial
K_{j}}
\end{equation*}
which can be calculated in a manner similar to that used to calculate$\
\langle S_{i}S_{j}\rangle $, and the subscript $\left( K\right) $ indicates
that the summation is over all possible couples of the spins belonging to
the original triple $i,k,m$.

\emph{\ }The fact that the correlation length of total density fluctuations
equals that of the ends/junctions fluctuations emphasizes the point that the
junctions induce an \textit{overall }effective attraction between the
monomers and \textit{not} only between the defects (i.e., ends and
junctions). In particular, the correlation length of junction-junction
correlations is the same as that of the \emph{monomer} density-density
correlations. The \emph{correlation length, }related to the\emph{\
thermodynamics }of the system\emph{,} is unrelated to the \textit{mean
distance\ }$\bar{l}$ between the junctions which reflects the \textit{%
structural} and topological properties.

\FRAME{ftbpFU}{3.2258in}{2.3765in}{0pt}{\Qcb{`Longitudinal' spin-spin
correlator measures the correlations between $any$ two chain ends. The
`transverse' correlator measures the correlation between the ends of the $%
same$ chain.}}{\Qlb{spinfig}}{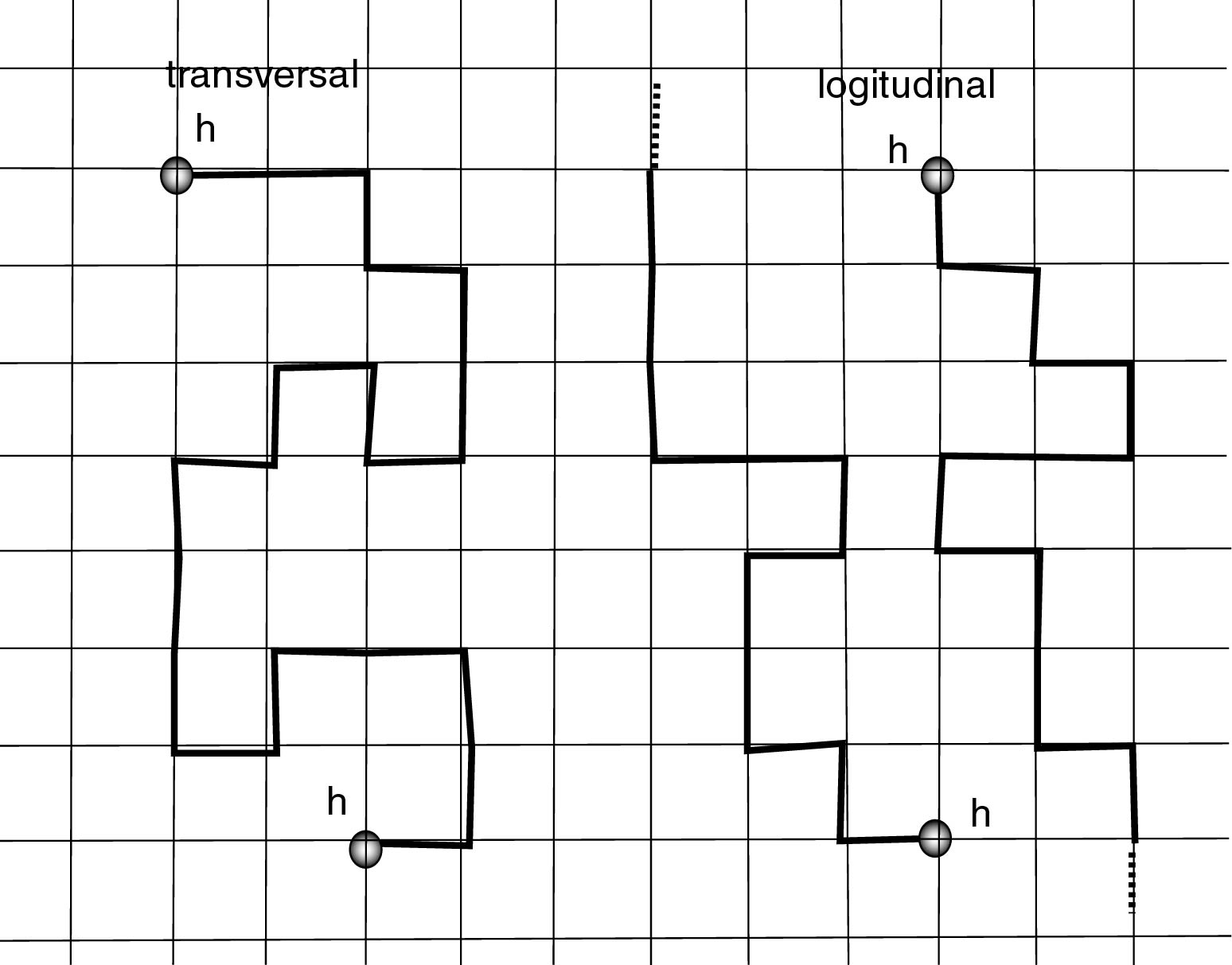}{\special{language "Scientific
Word";type "GRAPHIC";display "USEDEF";valid_file "F";width 3.2258in;height
2.3765in;depth 0pt;original-width 32.2497in;original-height
25.2707in;cropleft "0";croptop "1";cropright "1";cropbottom "0";filename
'L:/networks/translong.jpg';file-properties "NPEU";}}

\subsection{\protect\bigskip `Transverse' correlations: the structure of a
single aggregate and percolation\label{transversal}}

\bigskip As have been mentioned, the `transverse' correlation function
measures the autocorrelations (denoted by the subscript $auto)$ between the
ends of the \textit{same} cluster 
\begin{equation*}
\langle \phi _{e}(r_{i})\phi _{e}(r_{j})\rangle _{\text{auto}}=h^{2}\langle
S_{i}^{\bot }S_{j}^{\bot }\rangle
\end{equation*}
\emph{\ }Because $A^{\bot }=(1-\phi )$ is always positive, we can safely
expand the expression for the spin-spin correlator of Eq.$\left( \ref{cpa}%
\right) $ up to second order in the wavevector$\ p.$ Substituting the mean
field values of $J$ and $K$ from Eq.$\left( \ref{qJ}\right) $, one gets

\begin{equation*}
C_{\vec{p}}^{\bot }=\frac{(1-\phi )}{1-(1-\phi )A^{a}(Jg_{1}(\vec{p}%
)+6Ksg_{2}(\vec{p}))}\simeq \frac{1}{h_{0}/(2\phi )^{1/2}-3K_{0}\alpha
(2\phi )^{1/2}+bp^{2}}
\end{equation*}
where $b$ is a non-singular function of the parameters. In other words 
\begin{eqnarray*}
\langle \phi _{e}(\vec{p})\phi _{e}(\vec{p}^{\prime })\rangle _{\text{auto}}
&=&\frac{1}{(\phi _{e}-3\phi _{j})/2\phi +bp^{2}}\delta _{\vec{p},-\vec{p}%
^{\prime }} \\
\langle \phi _{e}(r)\phi _{e}(0)\rangle _{\text{auto}} &\simeq &\frac{1}{r}%
e^{-r/L}
\end{eqnarray*}
with $L=(2\phi /(\phi _{e}-3\phi _{j}))^{1/2}.$ This means that the
correlations between the ends belonging to the \textit{same} aggregate are
negligible for $r>>L$. For large separations $r$, the dominant contribution
to $\langle \phi _{e}(r)\phi _{e}(0)\rangle $ comes from the largest
cluster. In other words, the average size of the largest cluster is $\bar{R}%
\simeq L$%
\begin{equation}
\bar{R}\simeq \left( \frac{2\phi }{(\phi _{e}-3\phi _{j})}\right) ^{1/2}
\label{rbar}
\end{equation}
When $\phi _{j}=\frac{1}{3}\phi _{e}$, the mean size of the largest
diverges, and an infinite cluster is formed. This process is known as
percolation and the emergence of an infinite network at $\phi _{j}=\frac{1}{3%
}\phi _{e}$ agrees with other mean-field level studies of percolation \cite
{isaacson},\cite{florybook}. Substituting for $\phi _{e}$ and $\phi _{j}$
from Eq. (\ref{defsoffi}) we obtain the following equation for the
percolation line in the ($\phi ,T)$ plane (cf. Fig. \ref{fig1}): 
\begin{equation*}
\phi _{p}(T)=\frac{q}{6\alpha }e^{(\epsilon _{j}-\epsilon _{e})/T}
\end{equation*}
The length scale $L=\bar{R}$ has no thermodynamic meaning and purely
reflects a topological (infinite network vs. disconnected clusters)
property. At infinite temperature, the\emph{\ }junction-induced interaction
becomes irrelevant and $\phi _{p}$ is the percolation threshold for the
usual lattice percolation, studied in many classical works \cite{perc1}. The
value $\phi _{p}=\frac{q}{6\alpha }$ (equal to $\frac{1}{2(q-2)}$ for cubic
lattice)$,$ obtained here is not inconsistent with the usual result $\phi
_{p}=1/(q-1)$ for the site percolation; the difference stems from the fact
that in our model, there is percolation of two different objects: the
monomers and the junctions, which do not occupy the sites of the same
lattice; in our model, the monomers occupy the vertices of the lattice,
while the junctions occupy the interstitial positions.

Although the percolation transition, where an infinite cluster is formed is%
\textit{\ not} related directly to the junctions-ends transition discussed
in Sec. \ref{j-e}, the junction induced interaction influences the shape of
the percolation line at low temperatures\emph{,} in agreement with earlier
studies of the percolation in the interacting lattice gas \cite{perc1}. The
exact value of $\alpha ,$ and consequently of $\phi _{p},$ should not be
taken too seriously, because the lattice approximation is not adequate for
description of the system at the scale of an individual junction. Although,
the calculation shows that the percolation line cuts the coexistence line of
the phase transition on the left side of the critical point (cf. Fig. \ref
{fig1}), in real physical systems, the intersection point can be also to the
right of the critical point \cite{tanakaexp}. Rigid systems can have
effectively different $\alpha $, as discussed in Sec.\ref{rigid}.\emph{\ }%
Thus, depending on where the percolation line intersects the coexistence
curve, \ in general, the coexisting phases discussed in Sec. \ref{mf} can be
either two network phases, two phases of disconnected\emph{,} branched
clusters, or a network coexisting with dilute chains.

We now discuss the details of the cluster evolution as a function of monomer
density $\phi $. Recall that the mean distance between the defects (i.e.
junctions or ends) 
\begin{equation}
\bar{l}\simeq \frac{2\phi }{\phi _{e}+3\phi _{j}}  \label{lbartrans}
\end{equation}
is a non-monotonic function but has a maximum at $\phi _{e}=3\phi _{j}$; 
\emph{this is precisely the condition }that determines the \textit{%
percolation threshold}. Noting that Eq.$\left( \ref{rbar}\right) $ implies
that the cluster size increases \textit{monotonically} with increasing
density $\phi ,$ we conclude that there are two competing processes in
cluster formation: the lateral growth and the 'filling' in from inside. As
the monomer density is increased, for small densities up to the percolation
threshold, $\phi =\phi _{p}(T),$ the clusters become larger but more sparse.
At $\phi _{p}(T)$, an infinite cluster is formed and the process is
reversed: clusters (including the infinite one) become denser, being
predominantly filled in from inside. Although both below and above $\phi
_{p}(T)$ the lateral size of the clusters increases, the rate of \emph{%
lateral} growth with increasing monomer densities lower than the rate of 
\textit{internal }densification above $\phi _{p}(T).$

\section{Rigid and semiflexible chains\label{rigid}}

Solutions of rigid and semiflexible chains present two other broad classes
of systems which may show self-assembly and formation of ends and junctions,
in a manner similar to that of flexible chains as discussed up to now. As an
example, one can think about solutions or gels of actin$,$ a biopolymer
forming the cytoskeleton and involved in cell locomotion. Although it is
difficult to introduce bending rigidity into a lattice model such as ours,
the qualitative effects can be easily understood, at least at the mean field
level. For flexible chains, as in our model, the lattice size is usually
taken to be the persistence length (Kuhn segment for polymers), defining an
effective 'monomer' size, while for semi-flexible chains, the lattice size
would be the true size of molecular monomers, comprising the chains, the
persistence length being much larger than the lattice size.\ On very general
grounds, each defect (i.e. an end or a junction) contributes the energy $%
-k_{B}T$ to the free energy. Consequently, the defect-dependent part of the
free energy in a system where non-conserved defects are present is 
\begin{equation*}
F/T=-\phi _{\text{defects}}=-\phi _{e}-\phi _{j}
\end{equation*}
in accord with Eq. (\ref{Fdef}) \cite{tsvidip}. The difference between
flexible chains and rigid ones lies in the dependence of $\phi _{e}$ and $%
\phi _{j}$ on the total monomer density $\phi $, which can be understood on
the basis of the simple probabilistic argument outlined in Sec. \ref{j-e}.
Two ends can join to form a bond. In the case of flexible chains any two
ends which are neighbors on the lattice can form a bond, irrespective of the
orientation of the adjacent links (see Fig. \ref{juncfig}). In the case of
rigid rods, only rods that are co-linear, can coalesce. Therefore, taking
into account that the coalescence of two ends lowers the energy by an amount 
$2\epsilon _{e}$, so that the Boltzmann factor for the probability of
coalescence is proportional to $e^{2\epsilon _{e}},$ we find that in
equilibrium 
\begin{eqnarray*}
\phi &=&c\phi _{e}^{2}e^{2\epsilon _{e}} \\
c &=&q\text{ \ flexible chains} \\
c &=&1\text{ rigid rods}
\end{eqnarray*}
The same argument applies for a formation of a junction from three ends 
\begin{eqnarray*}
\phi _{j} &=&\alpha \phi _{e}^{3}e^{3\epsilon _{e}-\epsilon _{j}} \\
\alpha &=&q(q-2)/3\text{ \ flexible chains\emph{\ }(cubic lattice)} \\
\alpha &=&1\text{ rigid rods}
\end{eqnarray*}
In general, for semiflexible chains $1\leq c\leq q$; $\alpha \leq q(q-2)/3.$
In the continuum limit $q\rightarrow 4\pi $ for flexible chains. Therefore
the most general form of the mean field free energy of Eq.(\ref{FF}) can be
written:\ 
\begin{equation*}
F/T\simeq (1-\phi )\ln (1-\phi )-(\frac{\phi }{c})^{1/2}e^{-\epsilon
_{e}}-\alpha (\frac{\phi }{c})^{3/2}e^{-\epsilon _{j}}
\end{equation*}
with $c$ and $\alpha $ depending on rigidity of the chains and of the
junctions. In general, the degree of the flexibility of the junctions is not
necessarily equal to that of the chains themselves. The former may depend on
the microscopic properties of the cross-linker molecules as in the case of
gels or on the properties of the surfactant molecules in the case of
micelles and microemulsions, and other system-specific details.

The effect of the rigidity on the correlations is more difficult to treat
quantitatively in the present framework due to formal difficulty of
introducing the bending rigidity into the model. In particular, the
influence of branching on the nematic transition, present in the solutions
of rigid rods remains to be investigated. An heuristic theory \cite{drye}
might be more useful for this purpose.

\section{\protect\bigskip Discussion\label{discussion}}

We have presented a generic model of self-assembling chains which can branch
and form networks with branching points (junctions) of arbitrary
functionality. The model \textit{rigorously} maps the partition function of
a solution of branched, self-assembling, mutually avoiding chains onto that
of a Heisenberg magnet in the mathematical limit of zero spin components.
The model has been studied in the mean field approximation, which neglects
the presence of intracluster loops. It is found that despite the absence of
any \textit{specific interaction} between the chains, the presence of the
junctions induces an \textit{effective attraction} between the monomers,
which in the case of three-fold junctions leads to a first order reentrant
phase separation between a dilute phase consisting mainly of single chains,
and a dense network. The reason for this entropic transition lies in the
observation that although the translational entropy of the chains is lower
in the dense phase, the $total$ entropy is still higher due to the entropy
of the large number of self-assembling junctions in the dense phase. The
prerequisite for the phase separation is that the energy of a junction is
lower than the energy a free end: $\epsilon _{j}<\frac{1}{3}\epsilon _{e}$,
that is, the formation of junctions is energetically favorable (both $%
\epsilon _{e},\epsilon _{j}$ are considered positive and are measured
relative to the energy of a monomer in the middle of a chain). The model
then modified \ in order to study topological properties of the system at
the mean-field level. Independent of the junctions-ends transition, we
predict the percolation (connectivity) transition at which an infinite
network is formed. The percolation transition partially overlaps with the
junctions-ends transition (cf. Fig. \ref{fig1}). This transition takes place
at the point where the density of the free ends is three times the junctions
density, $\phi _{e}=3\phi _{j},$ which at the same time is the point where
the mean distance between the defects (i.e., junctions or ends), $\bar{l}$
attains its maximum. This result agrees with other mean field level studies
of percolation of tree-like clusters \cite{florybook},\cite{isaacson}. This
result means that up to the percolation transition the branched clusters
grow laterally, predominantly by addition of the new branches at the
periphery of a cluster, while at the percolation transition, an infinite
network is formed and the process is reversed: the clusters grow mainly due
to the filling of the internal mesh. The percolation transition is a
continuous, \textit{non thermodynamic }transition that describes a change in
the topology of the system but not a thermodynamic phase transition. Our
treatment which predicts both the thermodynamic phase equilibria as well as
the spatial correlations in the system allows us to treat both the phase
separation and the percolation threshold within the same framework, at least
at the mean field level.

The predicted density-density correlation function has a usual
Ornstein-Zernicke form at low monomer densities. The correlation length of
the density fluctuations $\xi $ is divergent at the spinodal line of the
junctions-ends transition and is a non-monotonic function of the monomer
density. $\xi $ attains its maximum around the point where the number of
ends equals the number of junctions, $\phi _{e}\approx \phi _{j}$. It is
important to emphasize that it is $\xi $ which is measured in actual
scattering experiments, and not other lengths of the problem, such as the
mean cluster radius $\bar{R}$ or the mean distance between the defects $\bar{%
l}$. The zero-wavevector scattering intensity, $S\left( 0\right) $ also has
a maximum as a function of $\phi $ along the same line, starting at the
critical point. This effect is a direct consequence of the first-order phase
separation between low and high density phases (cf. Fig. \ref{corrfig})

We also predict the emergence of the medium range correlations at high
monomer and junction densities, as reflected in the predicted peak in the
structure factor, signifying the onset of structural correlations in the
system (cf. Fig. \ref{corrfig}).

The theory presented here has many physical realizations. We now discuss the
implications of the generic results for several specific cases.

\subsection{Physical gels.}

Physical gels include a broad class of systems, consisting of long polymer
chains that are reversibly cross-linked. The cross-links can break and
reform under the influence of thermal fluctuations. Examples of these
materials are numerous, many of them of practical interest \cite{kumar}. The
chains that comprise the basis of the gel, can either be self-assembling or\
chemically bonded \cite{polymerization}. With increasing monomer density,
the gelation threshold is reached where a connected network (gel) is formed.
Despite intensive research in the past two decades, several features of the
physical gelation process are still under debate. One question concerns
whether the gelation transition (i.e., the transition from disconnected,
branched clusters to a macroscopic, connected network) is a structural
transition or a thermodynamic one. Another question of interest is:\ if the
gel transition is indeed a thermodynamic one, what is the order of the
transition \cite{tanakatheor,kumar}. Our results show that actually there
are \textit{two} independent transitions. The first one is the classical
percolation transition, located at the point where the density of the
cross-links is equal to one third of the density of the free ends in
agreement with classical result due to Flory \cite{florybook}. An infinite
connected cluster is formed at the percolation threshold. This transition is 
\textit{structural} and is not connected to any \textit{thermodynamic }%
singularity\textit{. }

For three-fold cross-links another, a \textit{thermodynamic} transition is
also present at small densities, and partially overlaps the percolation line
(see Fig. \ref{fig1}). This transition is the first order phase transition
due to inter-monomer attraction induced by the presence of the junctions and
not due to the aggregation of stickers (cross-link molecules). Our model
shows that the phase separation can occur under good solvent conditions, and
not only in $\Theta $-solvents. An advantage of the present model is that it
starts from a$\ rigorous$ description of a solution of self-avoiding
branched chains and does not involve any $ad$ $hoc$ hypotheses about
interactions between the monomers, cross-linkers and solvent molecules. An
interesting problem is the difference between$\ gelation$, when molecular
segments are allowed to polymerize and cross-link simultaneously in the
reaction bath, and $vulcanization$, when preexisting polymer chains are
cross-linked by irradiation, addition of linker molecules, or other means.
In addition to the fact that one expects large spatial inhomogeneities in
the case of vulcanization \cite{panykov1}, we argue that an additional major
difference is the functionality of the cross-links. In the case of gelation,
most of the junctions will be three-fold, simply due to the fact that the
collision of three segments is more probable than a collision of four. In
the case of vulcanization, on the contrary, the junctions will be four-fold,
being formed by cross-linking two pre-existing chains. So, in certain cases
gelation will be observed experimentally as a first-order thermodynamic
transition, while vulcanization is always continuous non-thermodynamic
transition. As a matter of fact, we have shown that the presence of
four-fold junctions is thermodynamically equivalent to decreasing the
quality of the solvent, in accord with the results obtained by other
theoretical methods \cite{panykov1} and Monte Carlo simulations \cite{kumar}.

The long-standing question about the presence of closed, intra-cluster loops
in the pre-gel clusters and in the gel phase, becomes irrelevant in the
present formulation. In our model, nothing prevents the intra-cluster loops
from forming and their number is determined by the distributions of ends and
junctions. However, we note that the question of the influence of the loops
on the thermodynamic and structural properties cannot be addressed properly
in the mean field approximation, used in the paper, because it effectively
disregards the loops.

Experimentally, the gelation transition is commonly observed by measuring a
viscosity increase due to the formation of the macroscopic network.
Therefore, the nature of the experimentally observed gelation transition
will depend on the path in the phase space along which the transition is
approached. For example, decreasing the temperature at constant monomer
density along the arrow $b$ of Fig.\ref{fig1} will be produce a continuous,
non-thermodynamic gelation transition at the point where line $b$ intersects
the percolation line. On the contrary, a temperature decrease along the
arrow $a$ will produce a first order phase separation between a phase of
dilute chains and a connected network (gel).

\subsection{Microemulsions}

Microemulsions consist of domains of two immiscible fluids (typically, water
and oil) stabilized by a surfactant \cite{gelbartbook,gommpershik}\emph{.}
One theoretical approach to their understanding is to map the oil-in-water
dispersion onto a lattice model in the following manner. When the
spontaneous curvature \cite{sambook} is small, the relevant length scale is
the persistence length of the surfactant film; this length is determined by
the bending rigidity and the thermal fluctuations \cite{taupin}. For systems
where the spontaneous curvature dominates (this occurs when the spontaneous
curvature is larger than the inverse of the persistence length), for a
certain range of spontaneous curvatures, one finds that\emph{\ }the oil or
water domains are either spheres or elongated tubes of radius $r\sim
c_{0}^{-1}$. For oil internal systems, the radius of the oil domains can
also be kept constant by maintaining fixed ratio of the surfactant density
to that of oil $\frac{\phi _{\text{surf}}}{\phi _{\text{oil}}}$ $($in a
water internal system it is given by the surfactant to water ratio). The
application of the general results of this paper to these cases requires
that one understands the mapping between the microemulsion and our lattice
model. Taking $r$ as the lattice constant, and considering an oil internal
system, the microemulsion can be mapped to a solution of self-assembling,
branched chains (oil domains$\rightarrow $`monomers'; oil+surfactant fraction%
$\rightarrow \phi $) as described by our model. Obviously, the analogy
breaks down at oil volume fractions close to one, because our model does not
allow the merging of a two elongated thin adjacent domains into a wider one.

Using this analogy, the evolution of spontaneous curvature dominated
microemulsions can be qualitatively described in the context of our model
and the results presented above, as follows. At low oil fraction, the
microemulsion consists of spherical droplets of radius $r\sim c_{0}^{-1}$.
At a certain oil fraction a `polymerization transition' \cite
{wheelersulf,tsvimicro} takes place and cylinders of radius $r$ are formed
which eventually branch. With increasing oil fraction at a certain oil
fraction $\phi _{p}(T)\simeq $const $\times e^{(\epsilon _{e}-\epsilon
_{j})/T},$ a percolating oil domain is formed, indicating the onset of the
bicontinuous microemulsion (cf. Sec. \ref{peak}). At low temperatures, this
transition is masked by a first order, \emph{phase separation }transition
due to junction-induced attraction between the oil domains (Sec. \ref{j-e}).
This first-order transition is present only if the branching points are
three-fold, which seems to be relevant to microemulsions. In addition to the
fact that three-fold branches are more probable statistically, the energy of
the three-fold junctions is lower than that of four- and higher fold
junctions, as can be shown from calculations of the bending energy of the
surfactant bilayers in the junction region \cite{tsvimicro}.

The structure factor (for scattering from the bulk water or oil domains --
i.e., bulk contrast experiments) has a simple Ornstein-Zernicke form in the
droplet microemulsion region (which includes also elongated but disconnected
cylindrical droplets) with a correlation length that diverges at the
spinodal line of the first-order transition. In the bicontinuous region, the
structure factor is an increasing function of the wavevector, as expected in
any high density system (cf. Sec.\ref{peak}). With increase in the number of
junction, either due to increase in the monomer density, or the temperature
concentration a peak appears in the structure factor, indicating the
emergence of medium range correlations between oil domains (cf. Fig. \ref
{corrfig}). The peak is predicted to appear whenever the condition $\phi
\gtrsim \frac{\alpha }{2q}e^{-2\epsilon _{3}/T}$ is satisfied. The peaked
structure factor is a distinctive characteristic of bicontinuous
microemulsions, as borne out by many experimental studies \cite
{gommpershik,strey}. It is important to emphasize, that our model predicts
that the peak occurs only when $\phi >\max [\frac{1}{2},\phi _{b}]$ which
means that, in principle there may be network microemulsions that \textit{do
not} show a peak in the structure factor ( the region between the
percolation line and vertical $\phi =1/2$ line Fig. \ref{corrfig}). The fact
that the peak appears only for $\phi >1/2$ should not be taken too
seriously. The exact value $\phi =\frac{1}{2}$ is most probably an artifact
of a lattice construction, reflecting the monomer-hole symmetry. More
important is the qualitative prediction that the peak in the structure
factor is a consequence of the correlation between the oil domains, induced
by the presence of the junctions.

\subsection{Wormlike micelles}

Surfactant molecules in aqueous solutions can form long, cylindrical
micelles in certain regions of the phase diagram. In the cylindrical phase,
each micelle consists of a large numbers of surfactant molecules. The
micelles are polydisperse in size, the equilibrium distribution being
determined by the interplay between the entropy (which favors small
micelles) and the `cap' energy, which is determined by the geometry of
packing of surfactant molecules at the ends of a micelle. Due to the bending
rigidity of the surfactant layer, the micelles are stiff up scales of order
of the \textit{persistence length }$l_{p}$ \cite{sambook}. On the other
hand, for length scales much larger than $l_{p},$ the micelles can be
considered as flexible chains. Therefore, for long enough micelles ($h_{0}/%
\sqrt{2\phi }\ll 1)$, the lattice constant of the equivalent Heisenberg
model can be taken to be equal to the persistence length $l_{p}.$ In the
opposite case of short chains, on can take a surfactant molecule as a
`monomer' size, with the\emph{\ }reservations described in Sec. \ref{rigid}.

It has been suggested on the basis of rheological, conductivity and
dielectric polarizability experiments, that the micelles can branch, and at
a certain point the system transforms to a connected micelle network. The
energy of the junctions and the ends in this system can be varied by
changing the salt concentration. Theoretically, it was proposed that the ` $%
n=0$' model might describe the formation of a network of wormlike micelles 
\cite{lekek,panizza}. Our results substantiate these suggestions and indeed
predict the formation of a connected micellar network, which can partially
overlap with the first order, phase separation phase transition.

Results of neutron scattering performed on wormlike micelles solutions show
that the correlation length, $\xi ,$ of the density fluctuations is a
non-monotonic function of the density in the proposed branched state \cite
{khatory}, which might be explained by the results of Sec. \ref{low}$.$
These experimental results were interpreted in Ref. \cite{khatory} in terms
of the mean micelle size (analogous to $\bar{l}$ in our formulation).
However, it is important to realize that, although both $\xi $ and $\bar{l}$
exhibit a maximum as a function of the density, it is $\xi $ which is
actually measured in scattering experiments, while $\bar{l}$ conveys purely
geometrical information, not directly measured in scattering experiments.
The correlation length $\xi $ has a maximum around $\phi _{e}=\phi _{j}$
which is a direct consequence of the junctions-ends transition, discussed in
Sec. \ref{j-e}. It should be noted that the maximum of $\xi $ is not
directly related to the network formation, which takes place at $\phi
_{e}=3\phi _{j}$, as well as $\bar{l}.$

\subsection{\protect\bigskip Dipolar and magnetic fluids and colloids}

Dipolar fluids or colloids consist of dipolar particles, bearing electric or
magnetic dipole. For, example ferrofluids. consist of magnetic (metallic)
particles immersed in an inert fluid. \ The molecules of real liquids, such
as acetone ($C_{3}H_{6}O)$ or dimethylsulfoxide ($C_{2}H_{3}SO)$ also bear
electronic dipolar moments. The interaction energy of two dipoles has a
minimum when the dipole moments are collinear \cite{jackson}. This means
that dipolar particles have a tendency to aggregate into chains \cite
{pincusdip}, so that all the dipole moments of the particles in the chain
are more or less colinear. However, it is important to realize that these
chains can also branch \cite{tsvidip}. The energy of the particles at the
branching point is obviously higher than in the middle of a chain and can be
calculated form electrostatics (or magnetostatics in the case of magnetic
dipoles) \cite{tsvidip}. Disregarding the \textit{long-range} part of the
dipole-dipole interactions, the system of both finite ended and branched
dipolar chains can be mapped to the model presented in this paper and our
predictions for the phase separation, scattering and percolation would
apply. This mapping is justified in dilute systems such as ferrofluids. or
magnetic colloids. In dense molecular liquids, it probably breaks down. In
order to understand the properties of molecular dipolar liquids, one may
have to take into account specific interactions between the molecules, such
as hydrogen bonding.

\subsection{Actin networks}

Actin is a self-assembling biopolymer, which supports the membranes of
biological cells. Sol-gel transition of actin solutions and networks,
induced by various proteins, (e.g., myosin) plays an important role in cell
movement and locomotion. The properties of actin networks $\ in$ $vivo$ have
been studied extensively by biochemical and molecular biology methods \cite
{molbio}. Recently, physical properties of actin solutions and gels have
been studied $in$ $vitro$, in rheological and scattering experiments \cite
{tempel}. The experimentally obtained phase diagram is strikingly similar to
Fig. \ref{fig1}, where at the percolation line there is a transition from
and \textit{entangled} \ network to a $connected$ `microgel' state. However,
it was observed that at low temperatures the dense phase at the coexisting
curve consists of 'bundles', where several parallel actin filaments are
tightly bound together by the cross-links \cite{tempel}.

Actin filaments are rigid, with the persistence length as high as ten
microns. However, the general results of this paper should apply, with the
reservations made in Sec. \ref{rigid}, which accounts for the similarity
between the predicted phase diagram and the experimentally found one. The
formation of the bundles is probably due to nematic interaction between
rigid filaments and cannot be captured in the present model. However, one
can estimate the condition for the bundles formation using a crude
qualitative argument. For this purpose, the network phase can be viewed as a
collection of rigid filaments of average length $\bar{l}\simeq e^{\epsilon
_{j}/T}\phi ^{-1/2}$ (cf. Sec.\ref{lbarsec}). The orientational order
appears in the solutions of rigid rods when $\phi \gtrsim $ const$/L$, which
means that for $\phi $ $\gtrsim e^{-2\epsilon _{j}/T}$ one might expect
appearance of the bundles.

\bigskip

\textbf{Acknowledgment \ }\textit{The authors thank R. Granek, J.-F. Joanny,
T. Lubensky, S. Panyukov, P. Pincus, Y. Rabin, E. Sackmann, T. Tlusty and T.
Witten for helpful discussions and the anonymous referee for illuminating
remarks. The support from ISF Center of \ Excellence on Self-Assembly, the
ACS and the donors of the Petroleum Research Fund and Schmidt-Minerva Center
is gratefully acknowledged}.

\section{\protect\bigskip Appendices}

\subsection{AppendixA: ` $n=0$' model\label{n=0app}}

The formal aspects of the `$n=0$' model have been studied extensively \cite
{degenbook,sarma,wheeler}. For the sake of completeness we review the
derivation here.\ Let us calculate the generating function Tr $e^{\vec{k}%
\vec{s}}$ for an $n$-component spin, $\vec{s}$, on a $d$-dimensional
lattice, where the averaging is over solid angle of the vector $\vec{s}$,
subject to condition $\sum_{\alpha }s_{\alpha }^{2}=n.$ 
\begin{equation*}
\text{Tr }e^{\vec{k}\vec{s}}=\frac{\int \prod_{\alpha }ds_{\alpha }\delta
(\sum_{\alpha }s_{\alpha }^{2}-n)e^{\vec{k}\cdot \vec{s}}}{\int
\prod_{\alpha }ds_{\alpha }\delta (\sum_{\alpha }s_{\alpha }^{2}-n)}=\frac{%
g_{n}(\vec{k})}{g_{n}(0)}
\end{equation*}
The cumulants of $\vec{s}$ can be obtained by differentiating the generating
function with respect to $k$. Now, supposing that $\vec{k}=(k,0,...,0)$%
\begin{eqnarray*}
&&g_{n}(k)=\int \prod_{\alpha }ds_{\alpha }\delta (\sum_{\alpha }s_{\alpha
}^{2}-n)e^{\vec{k}\cdot \vec{s}}=\int_{-\infty }^{\infty }d\omega e^{i\omega
n}\int ds_{1}e^{ks_{1}-is_{1}^{2}\omega }\int \prod_{\alpha
=2}^{n}ds_{\alpha }e^{-i\omega s_{1}^{2}}= \\
&=&\pi ^{n/2}\int_{-\infty }^{\infty }d\omega \frac{e^{i\omega n+\frac{k^{2}%
}{4i\omega }}}{(i\omega )^{n/2}}=\pi ^{n/2}i^{-n/2}[\int_{0}^{\infty
}d\omega \frac{e^{in(\omega -\frac{k^{2}}{4n\omega })}}{\omega ^{n/2}}%
+(-1)^{n/2}\int_{0}^{\infty }d\omega \frac{e^{-in(\omega -\frac{k^{2}}{%
4n\omega })}}{\omega ^{n/2}}
\end{eqnarray*}
The integrals in step two of the above equations, although formally
divergent, should be treated as generalized functions. The resulting
integrals are known \cite{ryzhik}. Putting $\nu =1-n/2,$ we find 
\begin{eqnarray*}
g_{n}(k) &=&\pi ^{n/2}i^{-n/2}2(\frac{\pi i}{2})(\frac{k}{2\sqrt{n}})^{\nu
}[H_{-\nu }^{\left( 1\right) }(ik\sqrt{n})+H_{-\nu }^{\left( 2\right) }(ik%
\sqrt{n})]= \\
&=&\pi ^{n/2}i^{-n/2}2\pi i(\frac{k}{2\sqrt{n}})^{\nu }e^{-i\frac{\nu \pi }{2%
}}I_{-\nu }(k\sqrt{n})=\pi ^{\frac{n}{2}+1}n^{\frac{n}{2}-1}\frac{I_{\frac{n%
}{2}-1}(k\sqrt{n})}{(k\sqrt{n}/2)^{\frac{n}{2}-1}}
\end{eqnarray*}
where $H_{p}^{(1)}(x),$ $H_{p}^{(2)}(x)$ are Hankel functions of the order $%
p $, and $I_{p}(x)$ is a modified Bessel function of the order $p$. Now, up
to second order 
\begin{equation*}
I_{p}(x)=(\frac{x}{2})^{p}(\frac{1}{\Gamma (p+1)}+\frac{1}{2\Gamma (p)}%
x^{2}+O(x^{4}))
\end{equation*}
Keeping in mind that in our case $p=\frac{n}{2}-1,$ it follows 
\begin{equation*}
\lim_{n\rightarrow 0}\text{Tr }e^{\vec{k}\vec{s}}=\lim_{n\rightarrow 0}\frac{%
g_{n}(k)}{g_{n}(0)}=1+\frac{1}{2}k^{2}
\end{equation*}
In particular, 
\begin{equation}
\text{Tr }s_{\alpha }^{m}=\frac{d^{m}}{dk^{m}}\text{{\LARGE [}}%
\lim_{n\rightarrow 0}\text{Tr }e^{\vec{k}\vec{s}}\text{{\LARGE ]}}%
_{k=0}=\delta _{m,2}\delta _{\alpha ,1}  \label{dergen}
\end{equation}
and 
\begin{equation*}
\text{Tr }s_{\alpha }s_{\beta }=\frac{\partial ^{2}}{\partial k_{\alpha
}\partial k_{\beta }}\text{{\LARGE [}}\lim_{n\rightarrow 0}\text{Tr }e^{\vec{%
k}\vec{s}}\text{{\LARGE ]}}_{k=0}=\delta _{\alpha ,1}\delta _{\beta ,1}
\end{equation*}

Due to the fact that the generating function has only quadratic terms in $k$%
, cumulants higher than second order vanish. This is because the higher
order cumulants are related to higher order derivatives of the generating
function with respect to $k$ and these all vanish, as seen from Eq. (\ref
{dergen})

\subsection{Appendix B: monomers, bonds and junctions on a lattice \label%
{alpha}}

\subsubsection{Relation between the densities}

The chains consist of monomers, occupying the vertices of the lattices,
connected by bonds. Each junction is not occupied by a monomer but is formed
by three unphysical 'bonds', connecting three adjacent monomers as
illustrated in \ Fig. \ref{juncfig}. The bonds forming the junction are not
counted by the model. The monomers comprising a chain fall into two
categories: internal ones, attached to two bonds, and the monomers at the
ends of the cluster, attached only to one bond. The total number of bonds
(including the unphysical ones) is therefore 
\begin{equation*}
N_{\text{bonds}}^{\text{total}}=\frac{1}{2}(2N_{i}+N_{e}+3N_{j})
\end{equation*}
where $N_{i}$ is the number of internal monomers, $N_{j}$ is the number of
junctions, and $N_{e}$ is the number of ends; the prefactor $\frac{1}{2}$
accounts for the double counting of \ each bond. The total number of
monomers is 
\begin{equation*}
N=N_{i}+N_{e}
\end{equation*}
and the number of physical bonds, connecting a pair of monomers, is the
total number of bonds less those unphysical bonds which are involved in
junctions 
\begin{equation*}
N_{\text{bonds}}=N_{\text{bonds}}^{\text{total}}-3N_{j}
\end{equation*}
Combining these equations together, one gets 
\begin{equation*}
N=N_{\text{bonds}}+\frac{1}{2}N_{e}+\frac{3}{2}N_{j}
\end{equation*}

\FRAME{ftbpFU}{3.3823in}{2.5071in}{0pt}{\Qcb{Two possible configurations of
the triplet $i,j,k$: a) the $j$ and $k$ sites are the nearest neighbors of
the site $i$; b) $k$ is the nearest neghbor of the site $i$ while $j$ is the
next nearest neighbor of $i$.}}{\Qlb{unphysjunc}}{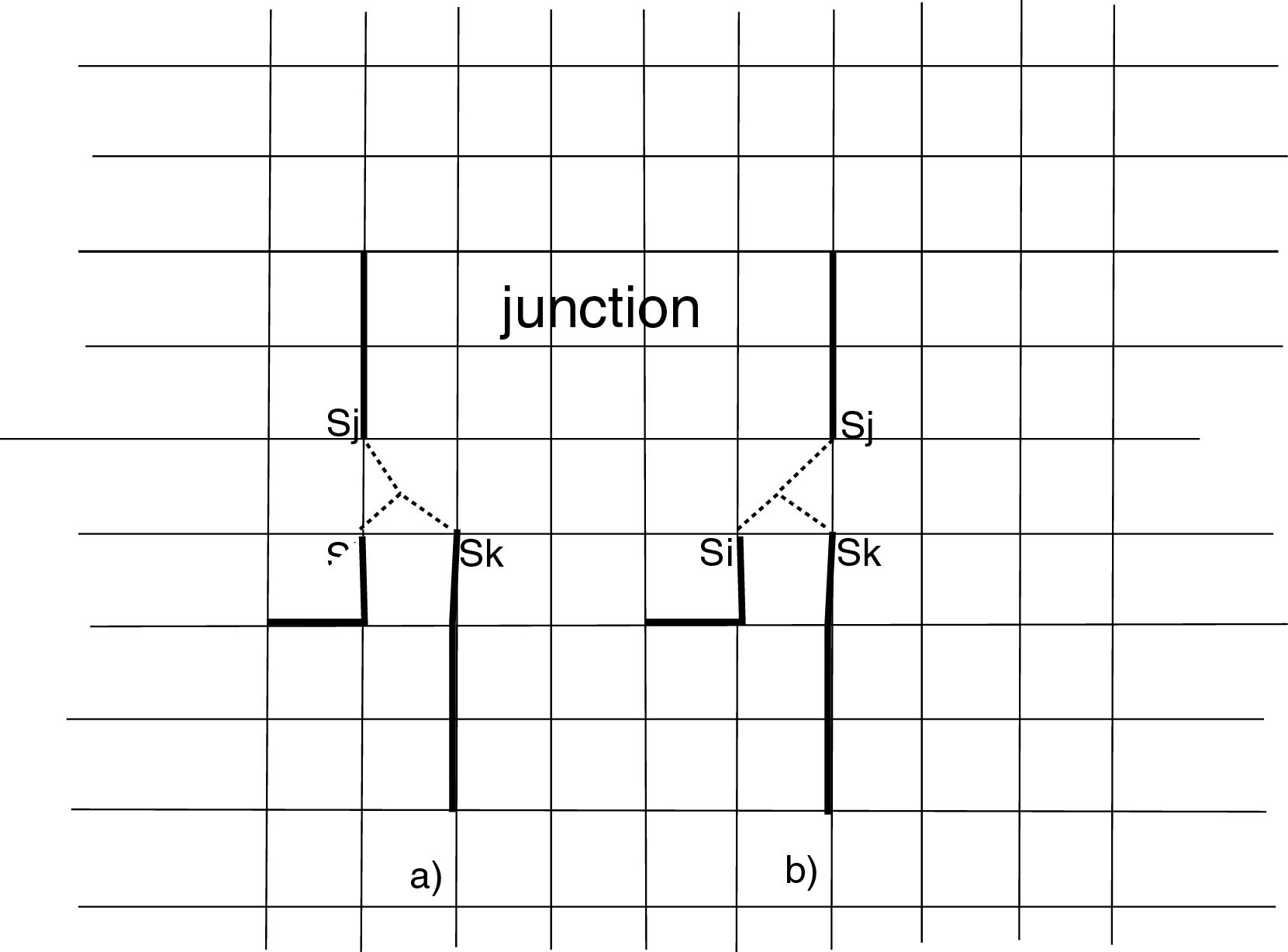}{\special%
{language "Scientific Word";type "GRAPHIC";maintain-aspect-ratio
TRUE;display "USEDEF";valid_file "F";width 3.3823in;height 2.5071in;depth
0pt;original-width 5.5348in;original-height 4.0906in;cropleft "0";croptop
"1";cropright "1";cropbottom "0";filename
'L:/networks/onejunction.BMP';file-properties "XNPEU";}}

\subsubsection{\protect\bigskip Value of $\protect\alpha $}

Here we present the considerations which lead to particular values of the
constants in mean-field equations of Sec. \ref{mf}. In our model we consider
each $site$ of a lattice occupied either by a monomer or solvent molecule.
Chain bonds are defined as the bond between two occupied neighbor sites. A
junction can be formed when three ends are neighbors on a lattice as shown
in Fig. \ref{juncfig}\emph{. } The sum in the three-spin term $%
\sum_{ijk}KS_{i}S_{j}S_{k}$ in Eq.\ (\ref{mf1}), \ accounting for the
junctions between the chains, is constrained over all \textit{distinct}
triplets of spins $i,j,k.$ \ Tthe three-spin term can be rewritten as $\frac{%
1}{3}\sum_{i}KS_{i}\sum_{jk}^{(K)}S_{j}S_{k}$ where $\sum_{i}$is now an 
\emph{unconstrained} summation over all cites and $\sum_{j,k}$ is \emph{%
unconstrained} sum over all possible pairs of sites $j,k$ which together
with $i$ form a junction $i,j,k$. In the mean field approximation this term
is fuirther transformed into $\frac{1}{3}\sum_{i}3KS_{i}\sum_{jk}^{(K)}%
\langle S_{j}\rangle \langle S_{k}\rangle $. On a hypercubic lattice, there
are two possible configurations of the triplet $i,j,k$. The first one, is
shown in Fig. \ref{unphysjunc}(a): both $j$ and $k$ are the nearest neigbors
of the site $i$. The second is realized when \ either $j$ or $k$ is a next
nearest neighbor of $i$, shown in Fig. \ref{unphysjunc}(b). A simple
calculation shows that for a hypercubic lattice of a coordination number $q$%
, the number of such pairs of the first kind is is $q(q-2)$ and the number
of the pairs fo the second kind is $\frac{1}{2}q\left( q-2\right) $.
Therefore, $\alpha =\frac{1}{3}(q(q-2)+\frac{1}{2}q(q-2))=\frac{1}{2}q(q-2)$%
. For a hexagonal close-packed structure or for the f.c.c. lattice, a
similar calculation gives $\alpha =q$ $(\ref{hexfig}$(a)). However, the
exact numerical value of $\alpha $ is an artefact of the lattice
construction and should be thought of as a phenomenological parameter
reflecting the microscopic features of a junction, along with the junction
energy $\epsilon _{j}$. A particular value of $\alpha $ affects only the
numerical values of the critical temperature and density but not the overall
qualitative behavior. It can be adjusted to reflect \ physically allowable
junction configurations.\ For example, three neigboring chains, coming
perpendicular to the plane of a triplet $\ i,j,k$ and connected by it, can
hardly be interpreted as a junction in most of the experimental systems. For
convenience, we have used $\alpha =\frac{1}{3}q(q-2)$ in calculating the
phase diagram and the structure factor. \FRAME{ftbpFU}{2.9905in}{2.1594in}{%
0pt}{\Qcb{Two possible alternative definitions of a junction. $a)$ shows a
junction on a hexagonal lattice$,$ $\protect\alpha =q/2$; \ $b)$ shows a
larger junction, coupling the sites separated by two lattice constants.}}{%
\Qlb{hexfig}}{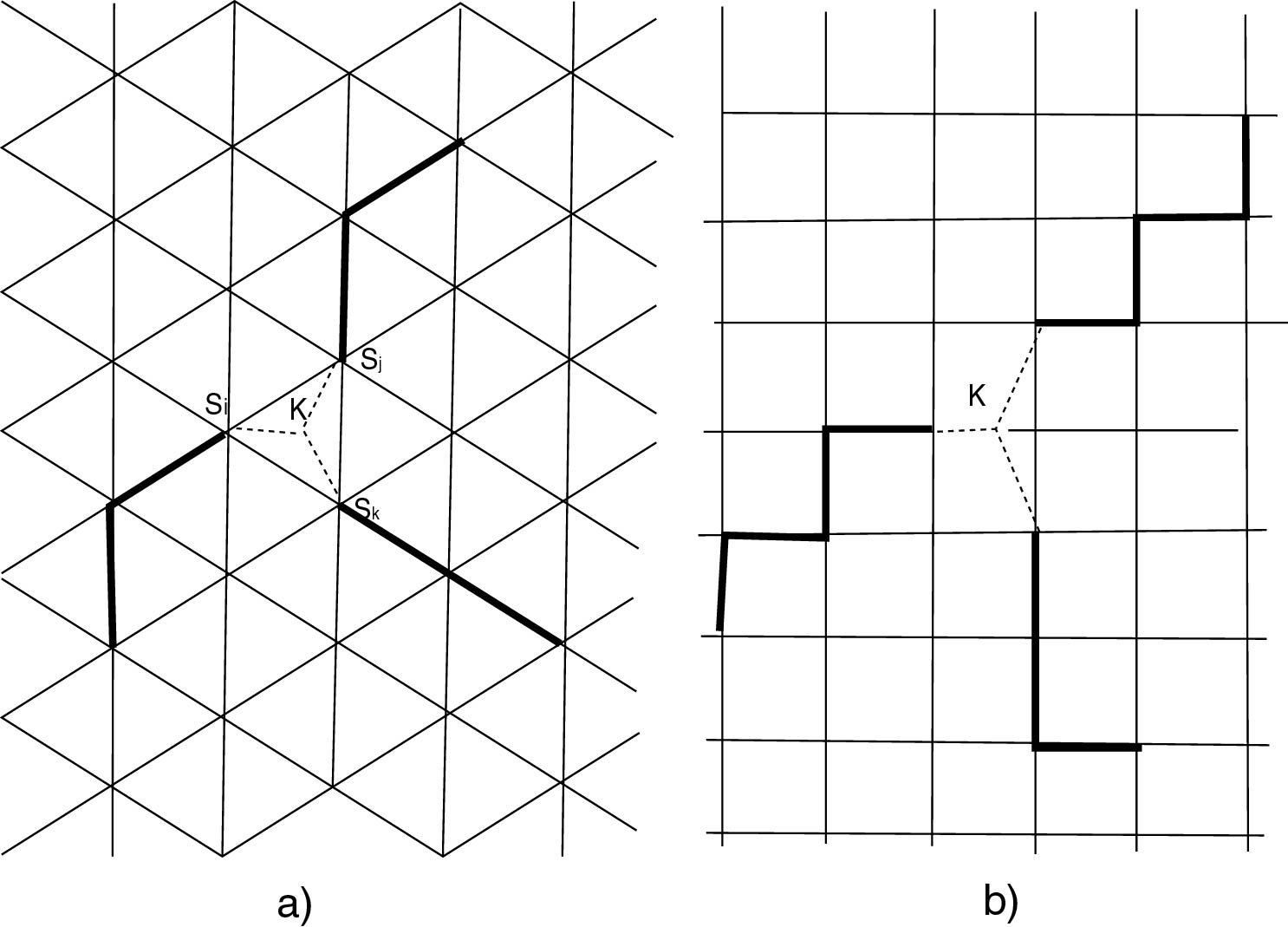}{\special{language "Scientific Word";type
"GRAPHIC";maintain-aspect-ratio TRUE;display "USEDEF";valid_file "F";width
2.9905in;height 2.1594in;depth 0pt;original-width 30.9482in;original-height
22.2706in;cropleft "0";croptop "1";cropright "1";cropbottom "0";filename
'L:/networks/hex.jpg';file-properties "NPEU";}}

\subsection{Appendix C: density-density correlations\label{dens-dens}}

\bigskip In this Appendix we show the derivation of the result (\ref{sij1})
of Sec. \ref{corr} for the density-density correlation function. From Eq.(%
\ref{jixisi})

\begin{eqnarray}
s_{i} &=&\frac{x_{i}}{1+\frac{1}{2}x_{i}^{2}}  \label{jixisiapp} \\
x_{i} &=&\sqrt{2\phi _{i}/(1-\phi _{i})};\text{ }\frac{\partial \ln x_{i}}{%
\partial \phi _{k}}=\delta _{ik}\frac{1}{2\phi _{i}(1-\phi _{i})}  \notag \\
s_{i} &=&\sqrt{2\phi _{i}(1-\phi _{i})};\text{ \ }\frac{\partial s_{i}}{%
\partial \phi _{k}}=\delta _{ik}\frac{1-2\phi _{i}}{\sqrt{2\phi _{i}(1-\phi
_{i})}}  \notag \\
J_{i} &=&\frac{x_{i}-h_{i}-K_{i}\sum_{j,k}^{\left( K\right) }s_{j}s_{k}}{%
\sum_{j}s_{j}}  \notag
\end{eqnarray}

Consequently, 
\begin{equation*}
\ln J_{i}=\ln x_{i}-\ln \sum_{j}s_{j}-\frac{h_{0}}{\sqrt{x_{i}\sum_{j}s_{j}}}%
-3K_{0}\frac{(\frac{1}{3}\sum^{\left( K\right) }s_{j}s_{k})x_{i}^{1/2}}{%
(\sum_{j}s_{j})^{3/2}}+O(K_{0}^{2},h_{0}^{2},K_{0}h_{0})
\end{equation*}

Now, 
\begin{eqnarray*}
\langle \delta \phi _{i}\delta \phi _{k}\rangle &=&\frac{\partial \ln J_{i}}{%
\partial \phi _{k}}=\frac{\partial \ln x_{i}}{\partial \phi _{k}}-\frac{%
\sum_{j}\frac{\partial s_{j}}{\partial \phi _{k}}}{\sum_{j}s_{j}}+\frac{h_{0}%
}{2(x_{i}\sum_{j}s_{j})^{3/2}}[\frac{\partial x_{i}}{\partial \phi _{k}}%
\sum_{j}s_{j}+x_{i}\sum_{j}\frac{\partial s_{j}}{\partial \phi _{k}}] \\
&&-3K_{0}(\frac{1}{3}\sum_{jp}^{\left( K\right) }s_{j}s_{p})[\frac{1}{%
2x_{i}^{1/2}(\sum s_{j})^{3/2}}\frac{\partial x_{i}}{\partial \phi _{k}}-%
\frac{3}{2}\frac{x_{i}^{1/2}}{(\sum_{j}s_{j})^{5/2}}\sum_{j}\frac{\partial
s_{j}}{\partial \phi _{k}}- \\
&&-3K_{0}\frac{x_{i}^{1/2}}{(\sum_{j}s_{j})^{3/2}}\frac{2}{3}%
(\sum_{jp}^{\left( K\right) }s_{p}\frac{\partial s_{j}}{\partial \phi _{k}})
\end{eqnarray*}
Together with Eq.$\left( \ref{jixisiapp}\right) $ this leads to Eq. (\ref
{sij1}) of Sec.\ref{densflucsec} .

\subsection{\protect\bigskip Appendix D: spin-spin correlations \label%
{spinspinapp}}

This Appendix shows the detailed calculation of the spin-spin correlation
function of Sec. \ref{spinspin}. In principle, the scalar products of any
two spins in Eq. (\ref{h3}) contain both longitudinal component, $S_{1}$,
parallel to the direction of the field $\vec{h}$, and a 'transverse' one, $%
\vec{S}_{\bot }$. For the two spin term it gives 
\begin{equation}
J\vec{S}_{i}\cdot \vec{S}_{j}=J\left( S_{1,i}S_{1,j}+\vec{S}_{i,\bot }\cdot 
\vec{S}_{j,\bot }\right)  \label{trans}
\end{equation}
and the three-spin term is chosen to be 
\begin{equation}
K\sum_{ijk}\left( S_{1,i}S_{1,j}S_{1,k}+\frac{\beta }{3}\left[ S_{1,i}\left( 
\vec{S}_{\bot ,i}\cdot \vec{S}_{\bot ,k}\right) +S_{1,j}\left( \vec{S}_{\bot
,i}\cdot \vec{S}_{\bot ,k}\right) +S_{1,k}\left( \vec{S}_{\bot ,i}\cdot \vec{%
S}_{\bot ,j}\right) \right] \right)  \label{transbeta}
\end{equation}
The average value of the `transverse' component, $\langle \vec{S}_{\bot
}\rangle ,$ is equal to zero, because the average magnetization is aligned
along the field $\vec{h}$. Consequently, it has no consequence for the
thermodynamic properties, and was neglected in the calculation of the free
energy in the Sec. \ref{mf}. However, its $fluctuations$ around its average
of zero, are not zero, and convey physically important information. Neither $%
S_{1}$ nor $\vec{S}_{\bot }$ has any $direct$ physical meaning. However,
physically relevant information can be extracted from the knowledge of their
correlation functions.

Consider the correlation function $\langle S_{1,i}S_{1,j}\rangle $ between
the 'longitudinal' components of two spins located at sites $i$ and $j$. As
follows from the discussion of the Secs. \ref{n=0},\ref{juncn0} the only
terms entering its expansion Tr $S_{1,i}S_{1,j}$ e$^{-H\left\{ S_{k}\right\}
/T}$ are the same as of Eq. (\ref{juncexpansion}) with $S_{1,i}S_{1,j}$
replacing the $h$-terms $hS_{i}$ and $hS_{j.}$ This means that the
correlator $\langle S_{1,i}S_{1,j}\rangle $ counts all the configurations
where there are chain ends at the point $i$ $and$ at the point $j$. In other
words, \ $\langle S_{1,i}S_{1,j}\rangle $ measures the correlations between $%
\ any$ two ends. Recalling that the ends density is $\phi
_{e}(r_{i})=h_{i}S_{1,i}$, and subtracting the $\delta _{ij}$ contribution
corresponding to the correlation of an end with itself, we arrive at Eqs. (%
\ref{endsr})(\ref{endsp}), which can also be obtained directly from the
relation 
\begin{equation*}
\langle \delta \phi _{e}\left( r_{i}\right) \delta \phi _{e}\left(
r_{j}\right) \rangle =\frac{\partial \phi _{e}(r_{i})}{\partial h_{j}}
\end{equation*}
\ \ To calculate the `transverse' correlation function, $\langle S_{i,\bot
}S_{j,\bot }\rangle $, where now $S_{\bot }$ signifies a component of $\vec{S%
}$ in one of the $(n-1)$ 'transverse' directions, one must focus on the
second term of Eq.$\left( \ref{trans}\right) .$ Following the arguments of
Sec. \ref{juncn0}$,$ the only terms which enter into expansion of $\langle
S_{i,\bot }S_{j,\bot }\rangle $ in the limit $n\rightarrow 0$ are of the
form 
\begin{eqnarray}
&&\text{S}_{i,\bot }[JS_{i,\bot }S_{m,\bot }\underset{\text{junction}}{%
JS_{k^{\prime },\bot }S_{k,\bot }\underbrace{[\frac{\beta }{3}KS_{m,\bot
}S_{n,1}S_{k,\bot }]}}JS_{n,1}S_{n^{\prime },1}...  \label{transterm} \\
&&...KS_{m^{\prime },1}S_{p,1}S_{q,1}.....JS_{j^{\prime },\bot }S_{j,\bot
}hS_{q,1}]\text{S}_{j,\bot }  \notag
\end{eqnarray}
that is, the clusters, consisting of the 'backbone' of the $S_{\bot }$'s
with branched `sidechains' of the $S_{1}$'s$,$ closed by $S_{i,\bot }$ and $%
S_{j,\bot }$. Inspection of possible terms of the form (\ref{transterm})
shows Eq.$\left( \ref{trans}\right) $that the sites $i$ and $j$ must belong
to the $same$ chain, because the Eqs.(\ref{trans},\ref{transbeta}) couple
either two transverse components or a longitudinal one with two transverse.
In other words, the `transverse' correlator measures the correlations
between any two ends belonging to a $same$ chain$.$ It is seen from equation
(\ref{transterm}) that the junctions enter the expansion of the transverse
correlator with the weight $\frac{\beta }{3}K.$ Since one wants to measure
the correlations between the ends of a cluster belonging to the expansion of
the original Hamiltonian ($\ref{h3})$, where all the junctions enter with
the same weight $K$, the value $\beta =3$ must be chosen. That is, the
relevant part of the three-spin term in the Hamiltonian $\left( \ref{h3}%
\right) $ has the form

\begin{equation*}
K\sum_{ijk}\left( S_{i}^{1}S_{j}^{1}S_{k}^{1}+S_{i}^{1}\left( S_{i}^{\bot
}S_{k}^{\bot }\right) +S_{j}^{1}\left( S_{i}^{\bot }S_{k}^{\bot }\right)
+S_{k}^{1}\left( S_{i}^{\bot }S_{j}^{\bot }\right) \right)
\end{equation*}

It is important to emphasize that the model given by Eq. $\left( \ref
{transbeta}\right) $ is suitable only for loopless clusters. Clusters
containing loops enter the expansion with wrong weights, due to summation
over $n-1$ 'transverse' components. However, the clusters with loops are
neglected in the mean field approximation and one can use the model given by
Eq.(\ref{transbeta}).

Performing the mean field approximation on these terms, as outlined in Sec. 
\ref{dens-dens} gives for the local spin averages :

\begin{eqnarray*}
s_{i,1} &=&\frac{\partial \ln Z}{\partial h_{i,1}}=\frac{x_{1}}{1+\frac{1}{2%
}[x_{1}^{2}+x_{\bot }^{2}]} \\
s_{i,\bot } &=&\frac{\partial \ln Z}{\partial h_{i,\bot }}=\frac{x_{\bot }}{%
1+\frac{1}{2}[x_{1}^{2}+x_{\bot }^{2}]}
\end{eqnarray*}
where 
\begin{eqnarray}
x_{1} &=&h_{i,1}+\sum_{j}Js_{j,1}+3(\frac{1}{3}\sum_{j,k}^{\left( K\right)
}Ks_{j,1}s_{k,1})  \label{x1xbot} \\
x_{\bot } &=&h_{i,\bot }+\sum_{j}Js_{j,\bot }+6(\frac{1}{3}%
\sum_{j,k}^{\left( K\right) }Ks_{j,\bot }s_{k,1})  \notag
\end{eqnarray}
with $\sum_{j}$ is the sum over nearest neighbors of the site $i$ and $%
\sum_{j,k}^{\left( K\right) }$ is the sum over the pairs of sites $j,k$
which together with $i$ make the junction triplet $i,j,k$ as in Sec. \ref
{spinspin}\emph{.} The coupling between the parallel and transverse
components of $s$ in the three-spin term in Eq.$\left( \ref{x1xbot}\right) $
has no consequence for the thermodynamic properties because the average
value of `transverse' component is zero: $\langle S_{\bot }\rangle =s_{\bot
}=0.$ The terms containing products $s_{i,\bot }s_{j,\bot }$ have been
omitted for the same reason. However, this coupling is important for
correlations, as we shall see. We thus define the spin-spin correlation
function $C_{ik}^{\left( a\right) }$, whose `longitudinal' component is
given by:

\begin{eqnarray}
C_{ik}^{(1)} &\equiv &\langle S_{i,1}S_{k,1}\rangle -s^{2}=\frac{\partial
s_{i,1}}{\partial h_{k,1}}=A^{(1)}[\delta _{ik}+J\sum_{j\in nn_{i}}\frac{%
\partial s_{j,1}}{\partial h_{k,1}}  \label{cik1} \\
&&+6K(\frac{1}{3}\sum_{m,j}^{\left( K\right) }s_{m,1}\frac{\partial s_{j,1}}{%
\partial h_{k,1}})]  \notag \\
\text{with }A^{(1)} &=&\frac{\partial }{\partial x_{1}}(\frac{x_{1}}{1+\frac{%
1}{2}[x_{1}^{2}+x_{\bot }^{2}]}){\Huge |}_{x_{\bot }=0}=\frac{1}{1+\frac{1}{2%
}x_{1}^{2}}-\frac{x_{1}^{2}}{(1+\frac{1}{2}x_{1}^{2})^{2}}  \notag
\end{eqnarray}
For transverse component, (bearing in mind that $\langle S_{\bot }\rangle
=s_{\bot }=0)$ we obtain: 
\begin{eqnarray}
C_{ik}^{(\bot )} &\equiv &\langle S_{i,\bot }S_{k,\bot }\rangle =\frac{%
\partial s_{i,\bot }}{\partial h_{k,\bot }}=A^{(\bot )}[\delta
_{ik}+J\sum_{j\in nn_{i}}\frac{\partial s_{j,\bot }}{\partial h_{k,\bot }}
\label{cikperp} \\
&&+6K(\frac{1}{3}\sum_{m,j}^{\left( K\right) }s_{m,1}\frac{\partial
s_{j,\bot }}{\partial h_{k,\bot }})]  \notag \\
A^{(\bot )} &=&\frac{\partial }{\partial x_{\bot }}(\frac{x_{\bot }}{1+\frac{%
1}{2}[x_{1}^{2}+x_{\bot }^{2}]}){\Huge |}_{x_{\bot }=0}=\frac{1}{1+\frac{1}{2%
}x_{1}^{2}}  \notag
\end{eqnarray}
Note that the coupling between $s_{\bot }$ and $s_{1}$ in Eq.$\left( \ref
{cikperp}\right) $ does contribute to the transverse spin-spin correlation
function. Substituting from Eq.(\ref{jixisi}) one finds 
\begin{eqnarray*}
A^{(1)} &=&(1-2\phi )(1-\phi ) \\
A^{\left( \bot \right) } &=&(1-\phi )
\end{eqnarray*}
Recalling that $\sum_{m,j}^{\left( K\right) }=\frac{2}{3}\frac{\alpha }{q}%
\sum_{nn_{i}}+\sum_{nnn_{i}}$, after Fourier transform $C_{ik}^{(a)}=\sum_{%
\vec{q}}e^{i\vec{p}(\vec{r}_{i}-\vec{r}_{k})}C_{\vec{p}}^{(a)}$ ($a=\left\{
1,\bot \right\} )$ the equations $\left( \ref{cik1}\right) $ and $\left( \ref
{cikperp}\right) $ yield the Eq. (\ref{cpa}) of Sec. \ref{spinspin}. It is
important to emphasize that the model given by Eq. $\left( \ref{transbeta}%
\right) $ is suitable only for loopless clusters. Clusters containing loops
enter the expansion with wrong weights, due to summation over $n-1$
'transverse' components. In principle, one can include higher order terms,
coupling different 'transverse' components. However, as seen from the
calculation above, in the mean field approximation, these terms $do$ $not$
contribute either to the free energy or to the spin-spin correlation
function. That is, the mean field results are independent of the precise way
in which the weights are assigned to clusters with loops. It follows that
the mean field approximation disregards the loops, which contain information
about long range intra-cluster correlations. Only the tree-like clusters
give contribution to the mean field approximation. It is also possible to
show by direct calculation from the rigorous Lubensky-Isaacson model that in
the mean field approximation the number of loops vanishes.

\subsection{Appendix E: Exact model for percolation}

In order extract the information about the topological structure of the
system, while retaining the exact correspondence between the expansion terms
of the spin model, and the equilibrium branched clusters, one has to resort
to tensor order parameter or, equivalently, to use several coupled $n=0$
fields. The presentation in this section follows closely the derivation of
Ref.\cite{isaacson} with the major distinction that the junctions are not
treated as point-like objects, which allows one to use a conventional
normalization of the $n=0$ spins $\vec{S}$. The Hamiltonian to be used is 
\begin{equation}
H_{T}=h\sum_{i}\sum_{\alpha =1}^{m}S_{1,i}^{\alpha }+J\sum_{\langle
i,j\rangle }\sum_{\alpha =1}^{m}\vec{S}_{i}^{\alpha }\vec{S}_{j}^{\alpha
}+K\sum_{\langle i,j,k\rangle }\sum_{\alpha =1}^{m}S_{1,i}^{\alpha
}S_{1,j}^{\alpha }S_{1,k}^{\alpha }
\end{equation}
where $\vec{S}^{\alpha }$'s are $m$ \ $n$-component vectors with the
normalization $\sum_{\alpha }(\vec{S}^{\alpha })^{2}=mn$ where $n\rightarrow
0$; $\alpha $ goes form $1$ to $\ m$. In this form, the Hamiltonian consists
of $m$ replicas of the Hamiltonian $\left( \ref{h3}\right) $, and there is
one to one correspondence between the terms in the expansion of Tr $%
e^{-H_{T}}$, and the equilibrium branched clusters. Due to presence of the
additional parameter, $m$, each cluster in the expansion acquires an
additional weight $m$. The partition function then reads 
\begin{equation*}
Z_{T}=\sum_{N_{j},N_{e},N_{b},N_{p}}J^{N_{b}}K^{N_{j}}h^{N_{e}}m^{N_{c}}%
\mathcal{N(}N_{j},N_{e},N_{b},N_{c})
\end{equation*}
where $N_{j},N_{e},N_{b},N_{c}$ are the number of junctions, ends, bonds,
and clusters, respectively. The parameter $m$ can be analytically continued
to non-integer values and has the meaning of the fugacity conjugate to the
total number of clusters in the system. For self-assembled branched
clusters, the relevant value is $m\rightarrow 1$, because the number of
clusters is not constrained \cite{isaacson}.

The Hamiltonian $\left( 1\right) $ can be rewritten, making the rotation in
the $m$-dimensional space of $\alpha $-components, with the use of a set of $%
m$-dimensional orthogonal vectors $\vec{v}_{l}$; $l=0,...,m-1$, with $\vec{v}%
_{0}=\frac{1}{\sqrt{m}}(1,1,....,1)$ and $\vec{v}_{l}\cdot \vec{v}%
_{l^{\prime }}=\delta _{ll^{\prime }}$ \cite{isaacson}. The new variables
are then defined as: 
\begin{eqnarray*}
\vec{S}^{0} &=&\sum_{\alpha }\vec{S}^{\alpha }v_{0}^{\alpha }=\frac{1}{\sqrt{%
m}}\sum_{\alpha }\vec{S}^{\alpha } \\
\vec{S}^{l} &=&\sum_{\alpha }\vec{S}^{\alpha }v_{l}^{\alpha }\text{ \ \ for
\ \ }l\neq 0
\end{eqnarray*}

Note that the normalization remains the same: $\sum_{l=0}^{m-1}$ ($%
S^{l})^{2}=mn$. The Hamiltonian expressed in the new variables then reads 
\begin{eqnarray}
H_{T} &=&h\sqrt{m}\sum_{i}\sum_{l=0}^{m-1}S_{i}^{l}+J\sum_{\langle
i,j\rangle }\sum_{l=0}^{m-1}S_{i}^{l}S_{j}^{l}+  \label{ex2} \\
&&+\frac{K}{\sqrt{m}}\sum_{\langle i,j,k\rangle }\left[
S_{i}^{0}S_{j}^{0}S_{k}^{0}+\left(
S_{i}^{0}\sum_{l=0}^{m-1}S_{j}^{l}S_{k}^{l}+\text{perm(}i,j,k\text{)}\right)
+\sum_{l\neq 0,l^{\prime }\neq 0,l^{\prime \prime }\neq 0}a_{ll^{\prime
}l^{\prime \prime }}S_{i}^{l}S_{j}^{l^{\prime }}S_{k}^{l^{\prime \prime }}%
\right]  \notag
\end{eqnarray}
where $a_{ll^{\prime }l^{\prime \prime }}$ are numerical coefficients,
depending on the value of $m$. It can be shown then that the correlation
function $\left\langle S_{i}^{0}S_{j}^{0}\right\rangle $ measures the
correlation between any two endpoints and the $\left\langle
S_{i}^{l}S_{j}^{l}\right\rangle $ measures the correlations between two
endpoints belonging to the same cluster \cite{isaacson}.

In the mean field approximation, the products of $\ S^{l}$ with $l\neq 0$ do
not contribute either to the free energy or to the two point correlation
function, because only $\left\langle S^{0}\right\rangle \neq 0$ (they do,
however, contribute to higher order correlation functions), and can be
neglected. In particular, the fact that the mean field two-point correlation
function is independent of the coefficients $a_{ll^{\prime }l^{\prime \prime
}}$ (i.e. , of the precise way in which weights are assigned to loops),
means that the mean field approximation takes into account only the loopless
graphs contributing to the two point correlation function. With this in
mind, the last term in Eq.(\ref{ex2}) vanishes, and in the limit $%
m\rightarrow 1$ the Hamiltonian $\left( 2\right) $ becomes formally
identical to the Hamiltonian (\ref{beta}) of Sec. \ref{spinspin}, with $%
S^{l} $ ($l\neq 0)$ taking place of a 'transverse' component $S_{\gamma }$, $%
\gamma \neq 1$.


\begin{thebibliography}{99}
\bibitem{polymerization}  In the case of 'living' self-assembled polymers, a
'polymerization' transition takes place at the very low densities, at which
long chains start to be formed as opposed to a solution of monomers (cf.
Ref. \cite{wheelersulf})
\end{thebibliography}
\end{document}